# Contextual stress and maternal sensitivity: a meta-analytic review of stress associations with the Maternal Behavior Q-Sort in observational studies

Anna T. Booth [a,b], Jacqui A. Macdonald (PhD) [a,b,c], George J. Youssef (PhD) [a,b]

Anna T. Booth: boan@deakin.edu.au
Jacqui A. Macdonald: jacqui.macdonald@deakin.edu.au
George J. Youssef: george.youssef@deakin.edu.au

[a]Deakin University, Centre for Social and Early Emotional Development, School of Psychology, Faculty of Health, Geelong, Australia.

Deakin University
221 Burwood Highway
Burwood VIC 3125
Australia

[b]Murdoch Childrens Research Institute, Population Studies of Adolescents, The Royal Children's Hospital Melbourne, Parkville, Australia.

Murdoch Childrens Research Institute
The Royal Children's Hospital
50 Flemington Rd.
Parkville VIC 3052
Australia

[c]The University of Melbourne, Department of Paediatrics, The Royal Children's Hospital Melbourne, Parkville, Australia

University of Melbourne Department of Paediatrics (RCH Academic Centre)
Level 2, West Building
The Royal Children's Hospital
50 Flemington Rd.
Parkville VIC 3052
Australia

**Corresponding author:**

Jacqui A. Macdonald
E: jacqui.macdonald@deakin.edu.au
T: +61 3 9246 8164

Deakin University
221 Burwood Highway
Burwood VIC 3125
Australia

Abstract

Maternal sensitivity is a modifiable determinant of infant attachment security and a precursor to optimal child development. Contextual stressors undermine sensitivity, but research was yet to be synthesized. We aimed to identify i) types of stress associations analyzed in studies of maternal sensitivity and ii) the strength of effects of various stress factors. A systematic search identified all studies that used the Maternal Behavior Q-Sort (MBQS) to code sensitivity in dyadic observations and that reported a coefficient for MBQS associations with contextual stress. Identified stressors cohered around three spheres: sociodemography (maternal education, family income, composite SES, maternal age and cohabitation status); parenting stress (perceived maternal stress related to parenting); and mental health (specifically maternal internalizing symptoms). Seven meta-analyses (combined *n*s range 223-1239) of a subset of 30 effects from 20 articles, and a multi-level meta-analysis (*N*=1324) assessed aggregated correlations with sensitivity. Significant mean effects emerged in expected directions, whereby all stress indicators were negatively associated with sensitivity. Small effects were found for associations with parenting stress (*r*=-0.13) and mental health indicators (*r*=-0.12). Generally moderate effects were found for associations with socio-demographic indicators (range *r*=-0.12 to *r*=0.32). Emerging findings support the proposition that in various contexts of stress, maternal sensitivity to infant needs can be undermined. Implications and research directions are discussed.

**Background**

Maternal sensitivity is a critical determinant of healthy infant development (Moran, Forbes, Evans, Tarabulsy, & Madigan, 2008). It predicts infant attachment security (Bailey, Redden, Pederson, & Moran, 2016; Bernier, Bélanger, Tarabulsy, Simard, & Carrier, 2014; De Wolff & van IJzendoorn, 1997; Moran, Pederson, Pettit, & Krupka, 1992; Pederson et al., 1990; Posada, Carbonell, Alzate, & Plata, 2004; Whipple, Bernier, & Mageau, 2011a), thereby laying foundations for socio-emotional competence across the life-course (Hazan & Shaver, 1994; Waters, Merrick, Treboux, Crowell, & Albersheim, 2000). Sensitivity is also relevant to a broad range of other offspring outcomes including executive function (Bernier, Carlson, Deschênes, & Matte-Gagné, 2012; Bernier, Carlson, & Whipple, 2010; Rochette & Bernier, 2014a); behavior problems (Bordeleau, Bernier, & Carrier, 2012b; Niccols & Feldman, 2006); sleep quality (Bordeleau, Bernier, & Carrier, 2012a); and, body mass index (Wendland et al., 2015; Wendland et al., 2014). Maternal sensitivity is a behavioral indicator of a caregiver's capacity to evaluate the type of care required by their child (Solomon & George, 1996) and involves the ability to perceive accurately and respond appropriately to the child's attachment-based signals (Ainsworth, Blehar, Waters, & Wall, 1978). Infants who experience sensitive caregiving develop confidence in their caregiver's emotional availability and responsiveness (Belsky & Fearon, 2002). When sensitivity is deficient, offspring in both infancy and preschool periods are at heightened risk of socio-emotional adjustment problems (Behrens, Parker, & Haltigan, 2011; De Wolff & van IJzendoorn, 1997; Posada et al., 2016). The predictive role of maternal sensitivity in child development is well understood (De Wolff & van IJzendoorn, 1997), but less is known about what predicts sensitivity.

To date, theorized predictive pathways to maternal sensitivity have centred on the role of adult attachment representations assessed by the Adult Attachment Interview (Bailey, Moran, Pederson, & Bento, 2007; Lindhiem, Bernard, & Dozier, 2011; Verhage et al., 2016; Ward & Carlson, 1995). Yet, meta-analysis indicates that as little as 12% of the variance in maternal sensitivity is explained by the AAI (van IJzendoorn, 1995), suggesting that more is unknown than known about the natural history of maternal sensitivity. Other evidence links child characteristics (Atkinson et al., 1999; Mills-Koonce et al., 2007) and biological or hormonal variations (Gonzalez, Jenkins, Steiner, & Fleming, 2012; van IJzendoorn, Bakermans-Kranenburg, & Mesman, 2008) to variation in maternal sensitivity. However, while associations are statistically significant, effect sizes are generally small. Important emerging evidence points to the potential impact of the social ecology on maternal sensitivity (Belsky & Fearon, 2002; Deschênes, Bernier, Jarry-Boileau, & St-Laurent, 2014; Lemelin,

Tarabulsy, & Provost, 2006; Logsdon et al., 2015; Pereira et al., 2012; Pianta, Sroufe, & Egeland, 1989; Posada et al., 1999; Rochette & Bernier, 2014b; Tarabulsy et al., 2005). In her seminal studies, Mary Ainsworth (1967) brought early attention to the relevance of socioeconomic conditions and stress when discussing the living conditions of mothers who were unable to provide high quality care. It has since been empirically recognized that prolonged or acute stress exposures in the family environment can indeed undermine sensitivity (Belsky & Fearon, 2002; Pianta et al., 1989). In the most recent revision of the *Handbook of Attachment*, Feeney and Woodhouse (2016) draw a link between parental sensitivity and the family stress model – a process model in which socioeconomic pressure undermines parenting quality (Conger & Donnellan, 2007). They direct attention to evidence of relationships between low socioeconomic status (SES) and parental sensitivity in studies by Mesman, van IJzendoorn, and Bakermans-Kranenburg (2012), Chaudhuri, Easterbrooks, and Davis (2009) and Yaman, Mesman, van IJzendoorn, Bakermans-Kranenburg, and Linting (2010) in which parental sensitivity tends to be lower in minority than majority groups as a consequence of low SES and stress related to ethnic minority status. However, no-one to date has systematically reviewed this literature, and Feeney and Woodhouse (2016) call for a deeper evidence-based understanding of ecological (or contextual) contributions to parental caregiving behavior.

To address this gap, the focus of the current review is the impact of contextual stress on maternal sensitivity. While we acknowledge the likelihood of transactional relationships between factors, for current purposes, we distinguish contextual contributions from representational, biological/hormonal and child contributions. We adopt a Bronfenbrennarian language, whereby individuals operate within and are influenced by their broader "socio-ecological systems" (Bronfenbrenner, 1986). Ecological risks associated with *parenting practices* are extensively documented (Conger et al., 1992, 1993; Kwon & Wickrama, 2014; Lee, Wickrama, & Simons, 2013; Newland, Crnic, Cox, & Mills-Koonce, 2013; White, Liu, Nair, & Tein, 2015). In contrast, less attention has been paid to the ecology of *caregiving sensitivity*. This distinction is important, given that parenting practices are understood within a social learning and behavioral modelling framework (Patock-Peckham, Cheong, Balhorn, & Nagoshi, 2001; Simons & et al., 1990), whereas caregiving sensitivity reflects an underlying affectional bond and dispositional response to the evolutionary goal of protecting and caring for a child (Solomon & George, 1996). Maternal sensitivity is primarily conceptualized within attachment and caregiving behavioral systems theory in which it is an indicator of the balance between self-regulation and regulatory support of a developing infant (Feeney & Woodhouse, 2016; Solomon & George, 1996).

Importantly, maternal sensitivity is modifiable through intervention, with outcomes as profound as enhancement of child attachment security (Bakermans-Kranenburg, van Ijzendoorn, & Juffer, 2003; Moss et al., 2011) and reductions in internalizing and externalizing problems (Moss et al., 2011). The potential for modification has prompted calls for investigations that identify barriers to sensitivity, and in particular consideration of stress or adversity within the familial ecology (Feeney & Woodhouse, 2016; Hyunjeong, Young-Joo, Hosihn, & Gyeong-Ae, 2008; Posada, 2013; Tarabulsy et al., 2005). Without prior aggregated assessment of effects, how relevant the social ecology is to parental sensitivity has not been well understood. Due to the expense of collecting the observational data used to code sensitivity, many studies in this area have drawn from small samples. While valuable, there are inherent weaknesses in smaller studies and findings must be interpreted with caution unless data are aggregated. Given the critical role of optimal sensitivity in a child's early life, these associations warrant meta-analytic investigation.

In the current review, we deconstruct representations of the stress ecology in extant literature. These include financial strain, reduced knowledge and opportunities that arise from low education and young parental age, and a diminished capacity to manage daily demands in a context of stress or mental health problems (Belsky, 1984; Berry & Jones, 1995; Conger & Donnellan, 2007). Our guiding proposition is that contextual stress is negatively linked to sensitive caregiving. When contextual stress-demands contemporaneously compete with caregiving goals, parental sensitivity is potentially challenged in a way that it is not for parents in more 'optimal' circumstances. Theoretical discussion of why contextual stress might be relevant to sensitive caregiving has been minimal (Feeney & Woodhouse, 2016; Solomon & George, 1996) and this review seeks to stimulate research interest in this area.

**Operationalization of Maternal Sensitivity**

We concentrate on a single operationalization of the sensitivity construct, which evolved from Mary Ainsworth's (1969; 1964; 1967) seminal naturalistic observations of maternal behavior in Uganda and Baltimore. The Maternal Behavior Q-Sort (MBQS) (Pederson et al., 1990) is an observational coding instrument that was developed with specific reference to the original Ainsworth sensitivity scale, which defines sensitivity as a caregiver's ability to perceive, accurately interpret and adequately respond to the needs of the infant (Ainsworth et al., 1978). Building upon this, the MBQS adapted Ainsworth's rather "global" impressions of sensitivity by very precisely quantifying what caregivers explicitly do in interactions with their infant (Posada, 2013). Using q-sort methodology, it weights the psychological significance of specific caregiving behaviors in a contingent, interactional context (Posada, 2013). Because of this particular methodology, and based on the especially extensive training process demanded of reliable coders, the MBQS provides highly standardized indices of sensitive caregiving that are theoretically translatable across contexts.

There is clear evidence that the MBQS is a strong predictor of infant attachment security (Behrens, Hart, & Parker, 2012; Moran et al., 1992; Pederson, Bailey, Tarabulsy, Bento, & Moran, 2014) and it has been found to explain more than 40% of the variance in attachment at the secure versus insecure classification level (Pederson et al., 2014). The MBQS can be reliably coded from a relatively short observation time (e.g., 20 minutes in a lab setting) in comparison with the original Ainsworth observations of up to several hours, yet generates an indicator of parental behavior that is functionally comparable to that determined by infant therapists (Moran et al., 1992). The MBQS is one of the most widely used sensitivity assessments in contemporary research: at least 90 published studies worldwide have empirically used a version of the MBQS across at least 13 countries, and there is demonstrated cross-cultural validity in MBQS sensitivity (Ekmekci et al., 2015; Emmen, Malda, Mesman, Ekmekci, & van Ijzendoorn, 2012; Posada et al., 2002).

In examining contextual stress associations with sensitivity, we elected to centre the investigation around a literature representing a single cohesive operationalization of maternal sensitivity. We aimed to provide a clear evidence base centred on a measure with demonstrated practicability in the research sphere and appealing translation potential. The MBQS exemplifies these qualities, as reflected by its widespread use in recent literature. It lends itself to high reliability and validity indices as it is often coded from videotaped interactions and subsequently rated by multiple coders who are trained under standardized requirements.

Our inquiry was focused on the potential for contextual pressure to undermine sensitivity at the behavioral level more readily than, for example, the representational level. The MBQS is especially appropriate as it was specifically designed to account for the caregiver's ability to attend to infant cues in a context of competing demands. This is important for the study of contextual stress associations because an ecology of stress will theoretically affect caregiver capacity to effectively divide attention and motivation across self-oriented and infant-oriented priorities.

**The Impact of Stress on Parental Sensitivity**

Developmental models in this area typically place maternal interactive behavior (e.g., sensitivity) in a mediating role between parental or contextual characteristics and child development (Tarabulsy et al., 2005). The assumption here is that parent-child interaction quality is the most relevant proximal factor for child development, and that for contextual factors to influence child development they must first impact on parental behavior (Tarabulsy et al., 2005). Stress or adversity in the social ecology might undermine sensitive parenting on a cognitive level because when negative emotion (e.g., stress or anxiety) is activated as a result of adversity, parents *appraise* parenting events differently (Dix, 1991). On an affective level, when stress demands are high, parents may be unable to activate the child-oriented *emotions* that motivate effective caregiving (Dix, 1991).

*Emotion regulation* is an underlying theme here and involves the processes by which individuals influence which emotions they experience, when they experience them, and how they express them (Gross, 1998). Emotion regulation underscores a person's deployment of attention, changes in cognitions, and modulation of responses (Gross, 1998). In the context of caregiving, emotion regulation involves parents' awareness and understanding of their own emotions; their appraisal of the effects their emotions will have if expressed; and their control of emotions (Dix, 1991). These factors map on to important features of the maternal sensitivity construct. A goal central to sensitive caregiving is co-regulation of child emotions (Feeney & Woodhouse, 2016; Jahromi, Putnam, & Stifter, 2004; Solomon & George, 1996) but a caregiver must first be able to self-regulate before they can regulate the emotions of the other (Feeney & Woodhouse, 2016). In a caregiving ecology in which there is stress or adversity, this emotion regulatory capacity might be challenged.

Sensitive caregiving promotes effectively balanced attention sharing between the demands of stress to the self (parent) and protection of the other (child) (Feeney & Woodhouse, 2016; Solomon & George, 1996). Attachment theory proposes that self-focused emotions (in the place of infant-focused emotions) interfere with sensitive caregiving because they can lead caregivers to withdraw from their child or respond intrusively (Feeney & Woodhouse, 2016). Caregiving is one of several *competing* motivational systems (e.g., attachment, affiliative and exploratory systems), and the balance that a parent strikes between these systems is affected by the demands made of them by their socio-ecological context (Dix, 1991; Solomon & George, 1996). In contexts of ecological stress where competing demands are heightened, emotional and attentional resources available for optimal caregiving might be compromised. Evidence shows that when mothers are depressed, for example, they experience fewer child-oriented emotions and concerns with a shift towards more self-oriented emotions (Dix, Gershoff, Meunier, & Miller, 2004); mothers with more infant-oriented goals demonstrate higher sensitivity in responding to infant distress than mothers with more self-oriented goals (Leerkes, 2010).

While it is noted that maternal sensitivity is conceptualized as a distinct construct from parenting, it is plausible that the ecological (or contextual) stress factors examined in the developmental and parenting literature will also impact on sensitivity (Feeney & Woodhouse, 2016). In particular, family stress model approaches to parenting behavior (Conger et al., 1992, 1993) highlight the links between socioeconomic adversity and parenting stress processes that increase parental emotional distress and undermine optimal child development (Conger & Donnellan, 2007). In this literature sociodemography encompasses low maternal and paternal education; low family income; low occupational status; young maternal age and single cohabitation status and is linked to family relationship problems and mental health risk (Conger et al., 1992, 1993; Conger & Donnellan, 2007; McLoyd, 1990; Newland et al., 2013). These inter-related contextual stress domains are also theoretically linked to caregiving sensitivity through an affective process model in which the level of balance achieved between motivational behavioral systems (e.g., attachment, caregiving, sexual etc.) is partly a function of socio-ecological stress demands (Dix, 1991; Dix et al., 2004). In optimal circumstances, these systems operate in harmony and are even complementary but under stress, they may compete for system resources. For example, a caregiver’s own adult attachment needs (a separate motivational system) might compete with the availability of their caregiving resources for infant needs. Co-existing systems may be concurrently or sequentially compromised by diversion of attention to contextual needs. With respect to the caregiving system in particular, low SES poses a risk to a parent’s capacity for contingent (i.e. sensitive) interactions with their child because of a preoccupation with environmental stressors (Crittenden & Bonvillian, 1984; Dix, 1991). However, empirical attention to the (often speculative) link between SES and caregiving sensitivity has been sparse (Tarabulsy et al., 2005). Measures of SES are often treated as controls rather than as having theoretical relevance in their own right (Conger & Donnellan, 2007). In this review we shift attention to SES as a central predictor of maternal sensitivity.

Maternal education too is a consistently important indicator of the parenting ecology (Bee et al., 1982; Brody & Flor, 1998; Cabrera, Fagan, Wight, & Schadler, 2011; Garai et al., 2009) and has been linked to the capacity to be emotionally available (Biringen et al., 2000). There is evidence that maternal education might uniquely be related to the capacity for sensitive parenting in ways that other socioeconomic factors are not. Education level is related to attentional capacity (Gómez-Pérez & Ostrosky-Solís, 2006) and attention to infant cues is essential for maternal sensitivity (Pederson, Moran, & Bento, 2015). The social context of low education might index competing attentional demands such as financial pressure (Conger, Conger, & Martin, 2010; Lochner & Moretti, 2004; Machin, Marie, & Vujić, 2011; Rauer, Karney, Garvan, & Hou, 2008). Also, high education status can act as a buffer against the effects of stress in the family (Almeida, 2005; Bee et al., 1982; Grzywacz, Almeida, Neupert, & Ettner, 2004), indicating a resilience to demands that might otherwise represent a vulnerability to the capacity for sensitive caregiving.

Similarly, young maternal age at child birth is likely to contribute to sensitivity for several reasons. Adolescent parenthood often happens in an adverse psychosocial climate that does not support an optimal parenting capacity (Barret & Robinson, 1981; Biello, Sipsma, & Kershaw, 2010; Neville & Parke, 1997; Tarabulsy et al., 2008). Young parental age is frequently accompanied by low SES (Bunting & McAuley, 2004; Jaffee, Caspi, Moffitt, Belsky, & Silva, 2001), heightened life stress, emotional unpreparedness to parent, and a lack of knowledge regarding child development (Barret & Robinson, 1981; Belsky, 1984; Biello et al., 2010; Elster, McAnarney, & Lamb, 1983; Neville & Parke, 1997). Differences in sensitivity according to maternal age might be a function of these related psychosocial stress factors and potentially by the demands of competing developmental roles for young mothers (the self: parent versus the other: child) (Dix, 1991).

Orientation to the child's needs is also threatened when mothers have mental health problems. A literature search (Hyunjeong et al., 2008) analysing the maternal sensitivity construct identified maternal depression, stress and anxiety as three key contextual 'negative affecting factors' that undermine sensitivity. Maternal mental illness is thought to divert attention from the child to the self, undermining the motivational structures that underlie supportive child-oriented parenting behavior (Dix, 1991; Dix et al., 2004). A parent's own perceptions of the stress that arises from the environmental demands of parenting is also relevant here (Anthony et al., 2005; Rodgers, 1998) and includes the appraisal of whether caregiving demands surpass available resources (Pereira et al., 2012).

**Aims and Scope of This Study**

Maternal sensitivity profoundly affects child development. Adult attachment is a key antecedent, but it only partly explains the variance in quality of maternal responsiveness. While stress in the family ecology has been widely studied, the impacts of contextual stress factors specifically on maternal sensitivity are not yet understood. We systematically review observational studies that report associations between contextual stress factors and maternal sensitivity coded using the Maternal Behavior Q-Sort (MBQS). A series of meta-analyses then explores the strength of emerging associations in order to understand the aggregated effects of various contextual factors that are theoretically relevant to caregiving behavior. We ground a timely theoretical discussion around the ecology of caregiving on the focal behavioral indicator of maternal sensitivity, and focus on a single exemplar operationalization of sensitivity at the interactive behavioral level. The objective of this review was to provide a theoretical framework and initial evidence base to justify broader inquiry. We aim to answer the following questions: in observational studies assessing sensitivity using the Maternal Behavior Q-Sort, (i) What contextual stress factors have been analysed in studies of maternal sensitivity?, and (ii) How strong are the associations between various contextual stress factors and maternal sensitivity?

# Method

## Measurement of Maternal Sensitivity

The Maternal Behavior Q Sort (MBQS) is an observational coding instrument based on Q-methodology used to assess maternal sensitivity in dyadic interactions. Versions of the MBQS have been developed for infants (Pederson et al., 1990) and more recently pre-schoolers (Pederson, Moran, & Bento, 2004). The MBQS can be used to code observations from both in-home and lab settings and is often applied to video-taped observational data.

Key behavioral indicators of maternal sensitivity coded by the MBQS include: accurate reading of and response to infant signals; synchronous interactions that revolve around infant tempo and mood-state; scaffolding of infant attention and exploration; attending to the infant in a context of competing demands; realistic expectations about infant emotion regulation; and, the ability to accept infant behavior even if it is not consistent with parental wishes (Pederson, Moran, & Bento, 1999).

The MBQS must be coded by trained observers. Training typically involves two to three days of structured observation, discussion and practice with an experienced expert. After further practice a coder should then demonstrate inter-rater reliability with a trained expert (typically consistency of 0.80 or higher).

Using the standard long-form MBQS (90 items) (Pederson et al., 1999; Pederson et al., 1990), observers sort items into nine groups of 10 items each based on how closely the items describe the parent under observation (a ranking system from *most unlike* to *most like* the parent). This generates a description of behavior (a sort) which is then correlated with a criterion sort that represents the ‘prototypically sensitive’ parent. The correlation between these two descriptions is the parent’s sensitivity score, which can vary from -1 (least sensitive) to +1 (prototypically sensitive). A short-form (Mini-MBQS; 25 items) (Moran, 2009; Tarabulsy et al., 2009) was adapted from the original so as to be less time-consuming yet well suited to providing a global measure of sensitivity. The MBQS-Mini includes items from the long-form that indicated the lowest and highest levels of sensitivity on the original 9-point scale, and items that were significantly associated with infant attachment classifications of secure or disorganized (Tarabulsy et al., 2009). Reliability and validity of the Mini-MBQS have been demonstrated against the long-form (Tarabulsy et al., 2009).

**Data Sources**

We conducted a systematic literature search of the following databases in August 2017 following PRISMA guidelines (Moher, Liberati, Tetzlaff, & Altman, 2009): PsycINFO, CINAHL Complete, MEDLINE Complete, EMBASE, SCOPUS, Cochrane Library and Google Scholar. Search terms were variations of ‘Maternal Behavior Q-Sort’ or ‘Maternal Behavior Q-Set’ with appropriate truncation and were designed to retrieve all studies that empirically used the MBQS (see supplementary Table S1). We did not include terms pertaining to contextual stress factors because these are variously conceptualized and often included in analyses as control variables. Screening all MBQS articles guaranteed inclusion of all relevant studies. Search criteria were restricted to English language, human, peer-reviewed articles where possible. We conducted the Google Scholar search without the peer-review restriction so as to assess grey literature. This search revealed that conference presentations, dissertations and other unpublished MBQS literature came almost exclusively from the same groups as those contributing to the peer-reviewed pool and this grey literature therefore did not provide a noteworthy additional contribution of target data. We did not apply publication date limits for any databases. A total of 823 articles were retrieved. After duplicate removal (*n*=132), 691 articles remained for screening.

**Study Selection**

We determined relevant articles by title, abstract and full text screening as necessary. We included articles for review if: i) the study used the MBQS to code sensitivity in an observational study; and, ii) a coefficient representing a contextual stress association with MBQS data was reported. Contextual stress factors were regarded as any possible indicators of the caregiving ecology theoretically implicated in sensitivity behavior because of their potential to exert a stress response. This excluded representational factors (adult attachment states of mind); infant characteristics (such as temperament); and biological markers (such as cortisol or genotype).

**Data Extraction**

We extracted and tabulated recruitment and sample characteristics from all articles meeting inclusion criteria (*N*=29). A standardized data table was used to extract and collate information from included studies. Based on the data collected from studies and through consensus by all authors, studies were organized into themes that reflected different indicators of the stress ecology: maternal age, maternal education, income, composite SES, absence of the infant's father from the home (cohabitation status of infant's parents), maternal internalizing symptoms and parenting stress.

**Screening for Meta-analysis Eligibility**

Much of the published literature comes from three key Canadian labs (Western Ontario; Montreal and Quebec City) and it was sometimes unclear whether separate articles were drawing from distinct or common samples. To clarify, we contacted researchers from these labs who independently confirmed cases of overlapping samples. Meta-analyses were conducted when there were two or more independent effect sizes per contextual stress variable.

We developed a decision hierarchy to filter studies for meta-analytic synthesis. If overlapping samples reported an association between MBQS data and the same stress variable in multiple publications, we excluded surplus associations in the following order: i) if the reported coefficient of the association was adjusted; then ii) if there was an analysis that drew from a larger sample size.

We extracted standard effect sizes (Pearson's *r*) from all but one of the eligible papers but converted effects to Pearson's *r* where necessary using standard formulas (Borenstein, Hedges, Higgins, & Rothstein, 2005). Associations were ineligible for meta-analysis if a single clear coefficient was not reported and if coefficients were reported for disaggregated MBQS subscales or factors only.

If multiple coefficients were reported in a single study for the same contextual stress variable, we retained the effect that was most consistent with other variables. For example, one study reported effects between maternal sensitivity and both unipolar and bipolar depression diagnoses, so we retained the unipolar effect because it was most consistent with other internalizing indicators across studies.

Data on sample size, effect size, effect type, *p* value, and study/sample information for conducting sensitivity analyses was independently extracted by two researchers for 100% of studies (AB 100%; JM 50% and GY 50%).

A total of 20 articles reporting 30 suitable effect sizes were eligible for aggregation in the series of meta-analyses, with data derived from 14 independent samples.

**Data Analysis**

Meta-analyses were conducted using the Metafor package v1.9.8 (Viechtbauer, 2010) in R software v3.3.1 (R Core Team, 2015) using inverse variance weighting (Viechtbauer, 2010) and included random effects to account for study heterogeneity. We tested for between-study heterogeneity using the $I^2$ statistic, which gives a percentage estimate of the amount of total variability in effect size estimates that can be attributed to heterogeneity among the true effects (Viechtbauer, 2010), although we caution over-interpretation of this statistic given the small number of effects in this study. The presence of publication bias was assessed by employing Egger's test for publication bias (Egger, Smith, & Phillips, 1997) where appropriate (Higgins & Green, 2011), and through visual inspection of the funnel plot from each meta-analysis.

We conducted sensitivity analyses to determine whether the overall meta-analytic effects were influenced by specific characteristics of the included studies or decisions made by the research team. In sensitivity analyses we retained the following conditions: mean infant age ≥6 months; unadjusted effect size; stress indicator measured as antecedent to maternal sensitivity; MBQS observation time ≥20 minutes; MBQS intra-class correlation (inter-rater reliability) ≥.80; MBQS 90-item (long form) used; and study did not draw from targeted risk sample. Sensitivity analyses could only be conducted if there were two or more studies remaining after exclusion according to these conditions (e.g., two studies in the 'maternal education' theme reported an inter-rater reliability at the desired level of ≥.80 and one did not). Given these requirements, the number of domains on which possible sensitivity analyses could be conducted was restricted.

Finally, we performed a meta-analysis in which the socioeconomic indicators (maternal education, family income and composite SES) were combined to compare the overall aggregated effect of the socioeconomic climate on sensitivity against the effects of its disaggregated components. This aggregation comprised 15 effects from 12 studies, representing associations between sensitivity and all socioeconomic indicators. To account for the clustered nature of the data (i.e., studies contributing multiple effects from the same sample) we used a robust variance meta-analysis with small sample adjustment (Hedges, Tipton, & Johnson, 2010; Tipton, 2015) using the Robumeta package v1.8 (Fisher, Tipton, & Hou, 2016) and examined the sensitivity of the obtained meta-analytic effect to variations in the assumed within sample correlation between different effects from the same sample.

**Interpretation of Effect Sizes**

When interpreting our findings, we focus on the meta-analytic effect size *r* of included associations for each aggregate analysis. We follow Cohen's (1992) guidelines for the interpretation of effect sizes whereby *r* of 0.1 = small, 0.3 = moderate and 0.5 = large.

# Results

## Identified Contextual Stress Factors

We identified a range of effects reported in the eligible literature that represented indicators of the social ecology theoretically relevant to caregiving behavior. These contextual stress factors cohered around three main clusters: indicators of the sociodemographic ecology (specifically maternal education, family income, composite SES, maternal age, and absence of the infant's father from the home), perceived parenting stress, and maternal internalizing symptoms (specifically unipolar depressive and anxiety symptoms).

The PRISMA chart (Figure 1) detailing the study selection process indicates the large number of articles (*n*=46) that were excluded based on the criterion that they did not report a coefficient for any association between sensitivity and contextual data. Many of the studies that reported eligible associations (*N*=29) did so in preliminary analyses only, typically controlling for stress indicators deemed to be possible confounding variables in analyses.

INSERT FIGURE 1 ABOUT HERE

## Characteristics of Included Studies

See Figure 1 for details of study selection. A total of 29 articles met inclusion criteria. Included studies were published between 1990 and 2017 and comprised a range of sample sizes (*n*=12 to *n*=379) and demographic types. The majority of studies drew from Canadian samples (*N*=21), with others from USA (*N*=2), Singapore (*N*=2), Colombia (*N*=1), China (*N*=1), South Africa (*N*=1) and Israel (*N*=1). See Table 1 for a comprehensive summary of recruitment and sample details including recruitment process, participation criteria, sample size, parent and infant gender composition, parent and infant age, and demographic information. Table 2 presents a summary of study predictors, outcomes, and findings of analyses that tested associations between maternal sensitivity and stress indicators.

Three studies sampled only at-risk participants (where risk was denoted by adolescent motherhood, low infant birth weight, or infant developmental delay). The samples in a further two studies comprised both high-risk and low-risk subgroups; one of these classified adolescent mothers as high risk compared to adult mothers, and one classified presence of depression/bipolar diagnosis as high risk compared to no diagnosis). The majority of studies drew from samples that were not characterized by any of the risk factors above.

INSERT TABLE 1 ABOUT HERE

INSERT TABLE 2 ABOUT HERE

**Measurement Methodology and Quality Assessment**

Table 3 provides a detailed overview of MBQS procedures adopted in each study including notes on the quality of coding and distribution of scores. Across studies there was heterogeneity in measurement quality of stress indicators (see Table 4) and in methodological use of the MBQS. There was also variability in how transparently studies acknowledged sample limitations and bias. Table 5 presents a summary of quality assessment of included studies against multiple assessment criteria.

INSERT TABLE 3 ABOUT HERE

INSERT TABLE 4 ABOUT HERE

INSERT TABLE 5 ABOUT HERE

**Contextual Associations with Maternal Sensitivity: Emerging Findings**

See Table 6 for a summary of findings on aggregated effect sizes, heterogeneity and publication bias. In each meta-analysis, a significant association emerged in the expected direction whereby higher levels of stress in the contextual domain were negatively correlated with sensitivity; and inversely, where lower levels of stress in the contextual domain were positively correlated with sensitivity.

To assess for publication bias, we examined funnel plots and performed Egger's tests on each aggregation. No statistically significant results were found. While we have presented results of publication bias testing in line with reporting conventions, these results should be interpreted with the caution that Egger's testing is not understood to be accurately interpretable in cases where there are substantially fewer than 10 studies per aggregation (Higgins & Green, 2011).

INSERT TABLE 6 ABOUT HERE

**Maternal education.** Around half (*n*=14) of all eligible studies (*N*=29) reported an association between maternal education and sensitivity. Based on a meta-analysis of nine effects (combined *n*=1059), maternal education was positively associated with maternal sensitivity, indicated by a small-to-moderate mean effect size (*r*=0.26, 0.18-0.34). This was significant at the *p*<.0001 level. Aggregated samples represent a diversity of populations (spanning low to high SES risk and various ethnicities). Moderate heterogeneity in effect sizes was found between studies ($I^2$ = 38%).

Where possible, we performed sensitivity analyses to examine the impact of a range of methodological conditions and sample characteristics on the effects of maternal education on sensitivity. See Table 7 for details. The aggregate effect was robust to the following conditions: where inter-rater reliability (IRR) among MBQS scores was <.80 and where the study drew from a targeted risk sample only. The effect of maternal education on sensitivity was slightly greater when studies that used a short form of the MBQS were excluded from the aggregate analysis (*r*=0.30, 0.19 - 0.41), when studies that coded from an observation period of <20 minutes were excluded (*r*=0.31, 0.22 - 0.40), and when studies with a sample of infants <6 months were excluded (*r*=0.32, 0.24 - 0.41). Using the long-form MBQS, longer observation periods, and samples of older infants produced slightly stronger effects that were associated with an additional 3-4% of the variance in maternal sensitivity. The use of a broader pool of descriptive items, and observations based on longer periods of time, may be important for clearly detecting any effects of stress on maternal behavior that are representative of typical dyadic interactions.

Effects were ineligible for inclusion in this meta-analysis where there was a larger overlapping sample contributing the same effect (*n*=4) (Bernier & Matte-Gagné, 2011; Bouvette-Turcot, Bernier, & Leblanc, 2017; Moran et al., 1992; Wen et al., 2017) and where a coefficient range only was reported (*n*=1) (Bordeleau et al., 2012a). Ineligible effect sizes (*n*=5) were comparable to those of included studies: *r*=0.25 (p<.01) (Wen et al., 2017); *r*=0.22 (p<.01) (Bouvette-Turcot et al., 2017); *r*=0.21 (*p*<.05) (Bernier & Matte-Gagné, 2011); *r*=0.24 (*NS*) (Moran et al., 1992); and a range of *r*=0.22 – *r*=0.34 (*p*<.05) (Bordeleau et al., 2012a).

**Family income.** Based on meta-analysis of four effects (combined *n*=490), family income was positively associated with maternal sensitivity, indicated by a small-to-moderate mean effect size (*r*=0.29, 0.19-0.39). This was significant at the *p*<.0001 level. Moderate heterogeneity in effect sizes was found between studies ($I^2$ =28%).

One association was ineligible for this meta-analysis as it was reported as a range only. This effect size was also small-to-moderate: *r*=0.22 – *r*=0.34 (*p*<.05) (Bordeleau et al., 2012a). The following two effects were ineligible for inclusion because there was a larger overlapping sample contributing the same effect: *r*=0.24 (p<.01) (Bouvette-Turcot et al., 2017); *r*=0.31 (p<.001) (Wen et al., 2017).

Where possible, we performed sensitivity analyses. The aggregate effect of income on maternal sensitivity was robust to the following conditions: where studies did not measure stress as an antecedent to sensitivity, and where inter-rater reliability (IRR) among MBQS scores was <.80. The effect of income on sensitivity was slightly greater when studies that coded from an observation period of <20 minutes were excluded (*r*=0.37, 0.25 - 0.50) and when studies with a sample of infants <6 months were excluded (*r*=0.35, 0.24 - 0.47).

**Composite SES.** Composite indicators of SES combined maternal education and family income (Rochette & Bernier, 2014b; Whipple et al., 2011a; Whipple, Bernier, & Mageau, 2011b), maternal and paternal education and family income (Bernier et al., 2014; Bernier, McMahon, & Perrier, 2017; Deschênes et al., 2014), and a composite index based on census data (Moran et al., 1992).

Based on meta-analysis of two effects (combined *n*=223), composite SES was positively associated with maternal sensitivity, indicated by a small-to-moderate mean effect size (*r*=0.29, 0.17-0.41). This was significant at the *p*<.0001 level.

Associations were ineligible for inclusion in this aggregation where there was a larger overlapping sample contributing the same effect (*n*=4) (Bernier et al., 2014; Deschênes et al., 2014; Whipple et al., 2011a, 2011b) and where they pertained to disaggregated MBQS factors only (*n*=6 in a single paper) (Rochette & Bernier, 2014b). Ineligible effect sizes ranged from negligible to moderate: *r*=0.37 (p<.001) (Deschênes et al., 2014); *r*=0.30 (*p*=.02) (Bernier et al., 2014); *r*=0.32 (*p*<.01) (Whipple et al., 2011a); *r*=0.29 (*p*<.05) (Whipple et al., 2011b); *r*=0.17 (*p*<.10); *r*=0.20 (*p*<.05); *r*=0.05 (*NS*); *r*=-0.03 (*NS*); *r*=0.23 (*p*<.05); and *r*=0.20 (*p*<.05) (Rochette & Bernier, 2014b).

Rochette and Bernier (2014b) report differential effects of composite SES against each of six MBQS factors. Significant, small-to-moderate effects were found for associations with each of the following factors: response to positive signals, response to distress, sensitivity/responsiveness and physical proximity. Negligible, non-significant effects were reported for positive affect sharing and hostility/rejection (Rochette & Bernier, 2014b).

**Combined socioeconomic indicators: maternal education; family income; composite SES.** When we combined the fifteen socio-demographic associations from twelve different studies using a robust variance meta-analysis with small sample adjustment, (combined $N$=1324), the aggregated effect of higher SES was small to moderate and positively associated with maternal sensitivity ($r$=0.26, 0.19-0.33; based on an assumed within cluster correlation of rho = .8). This was significant at the $p$=0.0002 level. Sensitivity analysis suggested that the meta-analytic effect was robust to a broad range of rho values.

Based on sensitivity analyses, the aggregate effect of the combined socioeconomic indicators on maternal sensitivity was robust to the following conditions: where socioeconomic indicators were not measured antecedent to MBQS sensitivity; where inter-rater reliability (IRR) among MBQS scores was <.80; and where the study drew from a targeted risk sample only. The effects of combined socioeconomic indicators were slightly greater when: effects drawn from studies using an observation time of <20 minutes to code the MBQS were excluded from the meta-analysis ($r$=0.33, 0.25 - 0.40); effects drawn from studies that used a short form of the MBQS were excluded ($r$=0.31, 0.23 - 0.39); and when studies with a sample of infants <6 months were excluded ($r$=0.31, 0.26 - 0.37).

**Maternal age.** Based on meta-analysis of four effects (combined $n$=264), maternal age was positively associated with maternal sensitivity, indicated by a moderate mean effect size ($r$=.32, 0.16-0.49). This was significant at the $p$=0.0001 level. Moderate heterogeneity in effect sizes was found between studies ($I^2$ = 41%) but our interpretation of this finding is cautious given the small number of effect sizes included.

**Absence of infant's father from the home.** Based on meta-analysis of two effects (combined $n$=391), absence of the infant's father from the home was negatively associated with maternal sensitivity, indicated by a small mean effect size ($r$=-0.12, -0.21- -0.02). This was significant at the $p$<.05 level.

Comparison of the effect size reported by Wade, Moore, Astington, Frampton, and Jenkins (2015) with the overall effect indicates that the very large Wade et al. (2015) sample ($n$=379) appears to have driven this aggregate result. The small and non-significant effect reported by Bigelow, Littlejohn, Bergman, and McDonald (2010), drawing from a very small sample ($n$=12), was of little impact to the overall effect in this model.

**Maternal internalizing symptoms.** Based on a meta-analysis of seven effects (combined $n$=1239), maternal internalizing symptoms were negatively associated with maternal sensitivity, indicated by a small mean effect size ($r$=-0.12, -0.18- -0.05). This was significant at the $p$<.0001 level.

A sensitivity analysis on the effect of timing of internalizing assessments relative to maternal sensitivity showed that the aggregate result was robust to the following conditions: where studies reported an adjusted effect size; where a short form of the MBQS was used; where the study drew from a targeted risk sample only; and where the mean infant age was <6 months. The effect of internalizing symptoms on sensitivity was slightly larger when internalizing symptomatology was measured antecedent to maternal sensitivity (*r*=-0.17, -0.30 - -0.04) and when studies using an observation time of <20 minutes to code the MBQS were excluded from the meta-analysis (*r*=-0.23, -0.37 - -0.09). Longer observation periods appear to be important for detecting additional variance (in this case 5%) in caregiving behavior related to internalizing symptoms, and using antecedent measurements of symptomatology may be optimal for identifying additional variance in sensitivity (in this case 2%) that could be related to more persistent internalizing symptoms.

A sensitivity analysis was conducted to examine whether the effect of internalizing symptoms on maternal sensitivity was robust to the inclusion of heterogeneous psychopathology types. We removed an indicator of anxiety symptomatology and an indicator of combined symptomatology (comprising anxiety, depression, irritability and cognitive disturbance), leaving only effects indicating depressive symptomatology (*n*=5). There was no change in overall effect (*r*=-0.10, -0.19 - -0.01).

Two associations were ineligible for aggregation because multiple coefficients for the same stress factor were reported in a single study. In the Logsdon et al. (2015) study, women with bipolar depression (the surplus effect) demonstrated lower sensitivity than woman with either unipolar or no depression (Logsdon et al., 2015). Secondly, in the Bailey et al. (2016) study there was a significant link between sensitivity and depressive symptoms at infant age 10 months (effect included) but not at age four months (the surplus effect).

Rifkin-Graboi et al. (2015) reported a marginally significant longitudinal association between maternal state anxiety at infant age three months and sensitivity at infant age six months. This study also tested associations between sensitivity and each of maternal prenatal state anxiety, prenatal trait anxiety, and postnatal trait anxiety but statistically significant effects were not found and effect sizes were not reported for these associations.

No eligible studies reported data on maternal externalizing symptoms.

**Parenting stress.** Parenting stress was measured by the Parenting Stress Index (PSI) (Abidin, 1990b) in all cases. For meta-analysis we were only able to aggregate effects corresponding to the parent domain of the PSI because there were insufficient samples reporting coefficients against other domains (child, interaction, life event and summed total domains) (Abidin, 1990a, 1995). The parent domain assesses general stresses associated with parenting such as feelings of parenting incompetence, partner conflict, perceived social support, and isolation due to the restrictions of parenting (Anthony et al., 2005; Pederson et al., 1990).

Based on meta-analysis of two effects (combined $n$=331), parenting stress was negatively associated with maternal sensitivity, indicated by a small mean effect size ($r$=-0.13, -0.23- -0.02). This was significant at the $p$<.05 level.

Seven associations were ineligible for aggregation because they indicated domains other than the parent domain. These effect sizes ranged from small to moderate in magnitude: $r$=-0.13 ($p$<.05); $r$=-0.16 ($p$<.05); $r$=-0.16 ($p$<.05) (Pereira et al., 2012); $r$=-0.36 ($p$<.05) (Pederson et al., 1990); $r$=-0.10 (*NS*); $r$=-0.15 ($p$<.10) (Tarabulsy et al., 2008) and $r$=-0.04 (Tarabulsy, Avgoustis, Phillips, Pederson, & Moran, 1997). The largest effect ($r$=-0.36, $p$<.05) (Pederson et al., 1990) was found for the association between maternal sensitivity and the child domain of the Parenting Stress Index (PSI), while smaller effects were found against the parent (and other) domains.

INSERT TABLE 7 ABOUT HERE

### Supplementary Results

Search terms used for systematic searching of databases are detailed in Supplementary Table S1. Supplementary Table S2 details our assessment of risk of bias in each study, used to inform sensitivity analyses on aggregated effects.

## Discussion

### General Findings

This review addresses an identified gap in understanding ecological contributions to maternal sensitivity. We found clear associations between increased levels of contextual stress or adversity and reduced levels of maternal sensitivity across three clusters: the sociodemographic ecology, perceived parenting stress, and maternal internalizing symptoms. Significant mean associations emerged in each of the seven meta-analyses. We report aggregated effects indicating that contextual factors are associated with between 1.4% (absence of father from the home) and 10.2% (maternal age) of the variance in maternal sensitivity. The importance of this is highlighted when compared to the seminal 12% variance in sensitivity explained by adult attachment representations (van IJzendoorn, 1995), to date the strongest theoretically causal predictor of maternal sensitivity. In the current study, small effects were found for associations with the parenting stress and mental health indicators, and small to moderate effects were found for associations with socio-demographic indicators. Effects ineligible for meta-analysis generally aligned with the directions of aggregated effects. Evidence should be interpreted with the caveat that few studies were available for each aggregation and at times large samples appeared to drive results. Notwithstanding, our findings are an important addition to an emerging evidence base and align with tenable theoretical explanations discussed below.

An explanation of current findings is that the identified stress indicators influence the capacity for sensitive caregiving when they denote an environment of adversity because this undermines the caregiver's ability to maintain attentional and emotional focus on their infant. The caregiver's own emotional resources are compromised due to heightened stress demands; they become more self-focused and less child-focused (Dix, 1991; Dix et al., 2004). Sensitive caregiving depends on effectively balanced attention sharing between the demands of stress to the parent and protection of the child (Feeney & Woodhouse, 2016; Solomon & George, 1996). Indeed, a key behavior evaluated on the MBQS is maternal skill in attending to the infant in a context of competing demands (Pederson et al., 1999).

## The Impact of Contextual Stress: An Emerging Evidence Base

**Maternal education.** Across studies, maternal education was one of the most commonly included contextual stress factors. Maternal education has been widely linked to quality of parenting behavior in the developmental literature (Bee et al., 1982; Brody & Flor, 1998; Cabrera et al., 2011; Garai et al., 2009) and has previously been associated with emotional availability (Biringen et al., 2000). Studies aggregated here span a range of education levels from <10 years' schooling (Bigelow et al., 2010) to samples in which more than half of mothers had at least a bachelor's degree (Wade et al., 2015; Xing, Zhou, Archer, Yue, & Wang, 2016). The samples that reported especially low mean education levels contributed strong effects to the aggregation (Bigelow et al., 2010; Tarabulsy et al., 2005). One possible explanation for this association is that education level is correlated with attentional capacity (Gómez-Pérez & Ostrosky-Solís, 2006) and attention to an infant's cues is a fundamental component of maternal sensitivity (Pederson et al., 2015). High education can enhance attentional capacity (Gómez-Pérez & Ostrosky-Solís, 2006), whereas the broader social context of low education potentially draws attention to competing problems such as relationship conflict or financial pressure (Conger et al., 2010; Lochner & Moretti, 2004; Machin et al., 2011; Rauer et al., 2008). Tarabulsy et al. (2005) postulate that the relationship between low education and low sensitivity in their study ($r$=0.39) might be explained by limited attentional resources available to adolescent mothers (Tarabulsy et al., 2005). In prior research, young mothers have been less aware of infant signals than adult mothers (Bailey, Waters, Pederson, & Moran, 1999). Maternal attention deficits have been associated with lower involvement with the child (Chronis-Tuscano et al., 2016), deficits in monitoring child behavior (Murray & Johnston, 2006), and lower maternal sensitivity after controlling for anxiety and infant activity (Semple, Mash, Ninowski, & Benzies, 2011).

In a study on a psychosocially vulnerable group of fairly low-educated mothers of low birth weight infants Bigelow et al. (2010) found a relatively large effect ($r$=0.45) between maternal education and sensitivity. It is possible that in young mothers, education level acts a proxy for maternal age, indicating a level of emotional maturity and plausibly affecting maternal sensitivity in this way. The effect did not adjust for maternal age; however, it was greater than the bivariate association between maternal age and sensitivity ($r$=0.37) (Bigelow et al., 2010). Findings from this comparatively low-educated sample suggest a potential increasing value of each additional year of education on sensitive caregiving. Education is understood to buffer against stress in the family (Bee et al., 1982): highly educated individuals report more daily stressors but are less reactive to them than less educated individuals (Almeida, 2005; Grzywacz et al., 2004). Low education might represent a particular vulnerability for the capacity to parent sensitively, whereby stressors in the family ecology are perceived as especially consuming relative to parents in higher education contexts.

The correlation between maternal education and sensitivity was slightly greater when studies that used a short form of the MBQS were excluded from the aggregate. The long-form MBQS might more comprehensively capture the range of possible deficits to sensitivity related to low education (such as impaired attention and response to infant signals), or the enhancements possibly related to high education (such as support for infant exploration and scaffolding of learning) due to the breadth of its larger overall pool of items. The short-form MBQS was designed to include a representative group of the broader pool of items that describe key sensitivity behaviors and is not thematically divergent from the original, but theoretically involves greater error at the item level which might contribute to smaller effects. The effect of education on sensitivity was also slightly greater when studies that had a sample of infants <6 months were excluded from the aggregate. Maternal education effects might be amplified in infants older than 6 months because deficits associated with low education will theoretically have had more time to erode contingent dyadic interaction patterns. Regarding high education effects, features of sensitivity that are likely informed by maternal attention and cognition (such as support for infant learning) might be more salient in dyads with older infants, when developmental goals require more support for exploration (Wade et al., 2015).

Two studies reported associations between *paternal education* and maternal sensitivity ($r$=0.29, p<.01) (Bernier, Jarry-Boileau, Tarabulsy, & Miljkovitch, 2010); ($r$=0.20, $p$<.01) (Bouvette-Turcot et al., 2017). While maternal and paternal education are often correlated (thus paternal education might indicate general family SES), this is not always the case (Breierova & Duflo, 2004). Further research is needed to identify whether maternal education has comparatively greater effects on sensitivity that might be explained specifically by resources available for responsive maternal interactions.

**Family income.** The overall finding that higher income was associated with higher maternal sensitivity aligns with a family stress model in which financial pressure undermines parenting quality (Conger et al., 1992, 1993; Conger & Donnellan, 2007). The heightened stress associated with a context of low income (Conger et al., 1992; Dix, 1991; Emmen et al., 2013; Newland et al., 2013) might deplete maternal emotional resources necessary for responding to infant needs. As such, sensitivity might be undermined though mediating pathways including perceived stress and mental ill-health that are associated with financial pressure (Emmen et al., 2013; Landers-Potts et al., 2015). More directly, competing stressors that might preoccupy a low-income mother (job-seeking; worrying over available resources; attending to other children and domestic duties alone if she is single) could impinge on the attention and time needed to adequately attune to infant needs.

Higher income likely provides contextual stability that reinforces effective maternal emotion regulation. From a hierarchy of needs perspective, the needs of the infant will best be met when their caregiver can operate in a secure environment and maintain an optimal balance between competing motivational systems (Dix, 1991; Leerkes, 2010; Solomon & George, 1996). In a context of stress related to low income, the caregiver's more proximate goal might be self-regulation (Feeney & Woodhouse, 2016; Solomon & George, 1996).

**Composite SES.** Composite SES (composite indicators variously comprising maternal and paternal education, family income and an index of census data) was one of the most commonly reported contextual factors across studies, but many effects could not be aggregated due to overlapping samples. Overall, higher composite SES was associated with higher sensitivity ($r$=0.29, $p$<.0001). Effect sizes from those studies ineligible for aggregation ($N$=9) ranged from negligible to moderate in strength.

The two samples contributing effects to this meta-analysis were distinct. First, the Bernier et al. (2017) study comprised intact families at low sociodemographic risk. This larger sample (*n*=204) drove the aggregate effect. In contrast, while very small (*n*=19), the Moran et al. (1992) study drew from a sample of wide SES range and with infants at-risk based on developmental delay (a biological rather than social factor). Moran et al. (1992) report a relatively low mean sensitivity score and suggest that this is explained by infants' inability to provide the level of transactional cues that non-delayed children would, with implications for their capacity to yield sensitive dyadic interaction (Moran et al., 1992).

In low SES groups, rates of disorganized child attachment (34%) are considerably higher than in general population groups (15%) (van IJzendoorn, Schuengel, & Bakermans–Kranenburg, 1999). Disorganized infant attachment involves the breakdown of a consistent strategy of emotion regulation (van IJzendoorn et al., 1999). A logical hypothesis would be that effects of SES disadvantage might profoundly undermine maternal sensitivity (a key predictor of child attachment security).

**Combined socioeconomic indicators: maternal education; family income; composite SES.** Given theoretical cohesion in the cluster of socioeconomic indicators (maternal education, family income and composite SES) in terms of their associations with maternal sensitivity, we aggregated these indicators in a robust variance meta-analysis to derive the overall effect of the socioeconomic climate on sensitivity. Given that effect sizes for subcomponent indicators were consistent in direction and magnitude (each demonstrated small to moderate positive associations with maternal sensitivity), it was reasonable to aggregate overall effects. Based on a large combined sample (*N*=1324), indicators of the aggregated socioeconomic ecology were associated with 6.7% of the variance in maternal sensitivity.

Sensitivity analyses indicated that an additional 3-4% of the variance in sensitivity was accounted for when effects from the following studies were removed: those that used an observation time of <20 minutes to code the MBQS; those that used a short form of the MBQS; and those that studied infants <6 months old. Further examination revealed that maternal education and family income effects from the Rifkin-Graboi et al. (2015) (*n*=271) study drove these findings. Associations in this study were weaker than in others but based on a large sample, effects reported by Rifkin-Graboi et al. (2015) received greater meta-analytic weighting.

Based on current aggregated findings, various types of socioeconomic stress appear to have similar associations with maternal sensitivity regardless of how they are indicated.

**Maternal age.** Overall, increasing maternal age was associated with higher maternal sensitivity ($r$=0.32, $p$=0.0001). This represents the largest effect size of all aggregated contextual indicators. The aggregation comprised a combined maternal age range of 16-45 years. A relatively strong effect contributed primarily by Tarabulsy et al. (2008) might be partly explained by effects of adversity in their high-risk subsample of adolescent mothers. Early parenthood often correlates with low SES (Bunting & McAuley, 2004; Hoff, Laursen, & Tardif, 2002; Jaffee et al., 2001; Lemelin et al., 2006), so maternal age effects in young mothers may well be a function of associated SES demands (low education and/or income).

There is likely variation in sensitive caregiving that is also a more direct function of maternal age. Early parenthood can be associated with emotional unpreparedness to parent and a lack of knowledge regarding child development (Barret & Robinson, 1981; Belsky, 1984; Biello et al., 2010; Elster et al., 1983; Neville & Parke, 1997). Adolescence is a period of rapid intellectual and emotional development, and even 18-year-old compared to 14-year-old mothers, for example, have likely undergone more substantial shifts in education, identity formation and social-cognitive development (Meeus, Van De Schoot, Keijsers, Schwartz, & Branje, 2010; Swanson, Spencer, & Petersen, 1998). Increased capacity for maternal sensitivity, based on heightened resources available for emotion co-regulation, is a plausible outcome of this. Possible heterogeneity in maternal age effects within ‘young’ groups of mothers might vary according to the extent to which maternal developmental goals are proximal at the expense of infant orientation. Adolescent mothers often have a high self-orientation and low child-orientation, and can have a low tolerance for negative emotion (Dix, 1991).

Maternal age might be differentially related to various domains of the sensitivity construct. Compared to young mothers, adult mothers are more aware of infant signals; more inclined to interact; and less ignoring and hostile/rejecting (Bailey et al., 1999). Further, the caregiving behavior of young mothers is much more heterogeneous than that of adult mothers. Three distinct styles of interaction quality have been found to vary among adolescent mothers: sensitivity-responsiveness, non-synchrony, and disengagement; and only one among adult mothers: sensitivity-responsiveness (Bailey et al., 2007; Bailey et al., 1999). These adolescent patterns of interaction are described as being paralleled in samples of depressed and abusive mothers, and might be more broadly common to mothers in high-risk environments (Bailey et al., 1999).

In future research, it will be important to examine the extent to which maternal age effects covary with associated socioeconomic adversity and how much variation might instead be explained by more specific age-related characteristics such as competing developmental goals or shifts in identity.

**Absence of infant's father from the home.** Absence of the infant's father from the home was associated with lower maternal sensitivity. Single parenting is a recognized correlate of low SES (Hoff et al., 2002) and might undermine sensitivity because maternal resources (financial and emotional) are likely to be compromised without partner assistance available for balancing child-rearing and domestic duties. Single mothers report more mental health problems and more daily hassles related to economic and family problems than mothers in two-parent families (Compas & Williams, 1990), and are more likely to suffer depression and chronic stress (Cairney, Boyle, Offord, & Racine, 2003).

The two samples in our aggregation differed dramatically in size and type. Wade et al. (2015) drew from a large sample ($n$=379) that included more intact families than the general population of the region. This bias might explain the small overall association with sensitivity, as cohabitation effects might be amplified in a more representative population including more single-parent families. Single parents struggling with competing stress demands might be less inclined to participate in research. Findings from the very small ($n$=12) Bigelow et al. (2010) sample contribute little to the aggregate effect size. In this sample, most (10 of 12) mothers were cohabiting with the father of the infant, and authors note that potential mediating or moderating effects of relationship quality, intimate partner violence or partner support were not examined. This is also the case for the effect reported by Wade et al. (2015). The presence of a supportive partner can promote optimal family functioning (Roye & Balk, 1996; Stapleton et al., 2012).

Effects included here that indicate only presence or absence of the infant's father might be rudimentary indicators of the impact of contextual adversity, as they do not account for whether the partner is supportive. Future studies should examine possible relationships in large, representative samples, or by drawing from multiple samples across risk and cultural profiles that allow for aggregation of specific effects.

**Maternal internalizing symptoms.** Maternal internalizing symptoms were one of the more commonly reported contextual factors across studies, and were associated with slightly lower maternal sensitivity overall. Mental illness theoretically diverts attention from the child to the self, and is proposed to destabilise the motivational structures that necessitate responsive infant-oriented caregiving (Dix, 1991; Dix et al., 2004). Supporting (cross-sectional) evidence indicates that increasing maternal depressive symptoms are linked to mothers having fewer child-oriented concerns and fewer child-oriented positive emotions than mothers with less depressive symptoms (Dix et al., 2004) as well as delayed and muted affective responses, impatience, and impaired communication skills with children (Lovejoy, Graczyk, O'Hare, & Neuman, 2000). Depressive symptomatology can disrupt emotion regulation processes which inform attention, cognition, and modulation of responses (Gross, 1998). Resulting cognitive deficits might affect features of sensitivity such as awareness of infant needs and response effectiveness. Affective deficits might undermine maternal delight, praise, and synchrony in interactions (which are contingent on caregiver ability to match infant mood-state).

A limitation of our meta-analysis was heterogeneity in the measurement of internalizing symptoms. Methodologies were as diverse as clinical diagnoses and self-report questionnaires. Self-report data can be confounded by bias related to social desirability effects (Chung & Monroe, 2003; Fisher & Katz, 2008; Van de Mortel, 2008). One study (Tarabulsy et al., 2005) used a general depressive symptom checklist, the CES-D, and did not find an effect between maternal depression and sensitivity. Authors suggest that this instrument may not be sensitive enough to adequately capture the nuances of symptomatology in specific populations such as adolescent mothers (Tarabulsy et al., 2005). We included indicators of anxiety in this aggregation, but there are logical reasons to argue differential effects of anxiety on maternal sensitivity. Anxious parents are potentially hypervigilant to infant threat, resulting in overprotection and intrusiveness. Alternatively, in a context of stress, anxious parents might exhibit hypervigilance to self-oriented threats that compete with infant needs. This may result in a failure to observe infant cues or mistiming and/or misdirection of responses to infant cues. However, when we examined the effects of depressive symptomatology only, there was no difference to the overall effect on sensitivity. It will be important for future studies to further examine possible differential effects of internalizing symptomatology. Broadly, these effects are expected to stem from a pervasive disruption to self-regulatory capacity whereby the caregiver is unable to regulate infant emotions in this context of depleted psychological resources (Dix, 1991; Dix et al., 2004; Feeney & Woodhouse, 2016).

A key finding within the broader review is that maternal sensitivity might be vulnerable to depression duration effects but robust to transient symptomatology. Bailey et al. (2016) found a relationship between sensitivity and maternal depressive symptoms at infant age 10 months but not 4 months. There is understood to be continuity in internalizing symptoms over time (Flett, Vredenburg, & Krames, 1997), so depressive symptoms reported at 10 months were likely longer lasting than symptoms reported at 4 months. Depressive symptoms might undermine sensitivity only in contexts of chronicity. Previous research in longitudinal case-control studies (Campbell, Cohn, & Meyers, 1995; Dannemiller, 1999) has reported similar findings. Alternatively, the additional demands likely experienced by the mother of a 10-month old infant compared to a 4-month old infant might expose underlying deficits in sensitivity.

We could not ascertain directionality of effects for these associations. While the assumption is that internalizing symptoms undermine sensitivity, the inverse is also possible. Insensitive mothers might be confronted by related infant behaviors that could undermine their sense of competence and reduce shared positive affect. In cross-sectional studies where the timing of symptom onset is unclear, we cannot clearly understand the direction of effects. A sensitivity analysis indicated that effects that used only antecedent measures of internalizing symptoms were not, however, significantly different from the overall effect (likely due to the small number of eligible studies).

**Parenting stress.** Our aggregation of two effects produced a small overall association between perceived parenting stress and lower maternal sensitivity. Parenting stress indicates a parent’s appraisal of whether caregiving demands surpass available resources (Pereira et al., 2012), which is likely the case in an overall context of adversity (Dix, 1991).

Effect sizes that were ineligible for aggregation ($N$=7) were all negative and mostly marginal or small with the exception of a moderate negative effect reported by Pederson et al. (1990) against the child domain of the PSI. This domain evaluates parental stress arising from child characteristics and encompasses factors such as the child’s demandingness and reinforcement of the parent’s behavior (Abidin, 1990b). The larger child domain effect suggests that parental perception of child characteristics might play a greater role in contingent caregiving than stress related to factors such as perceived parental incompetence, social support and partner conflict.

Again, cross-sectional analyses prevented conclusions about directionality in associations. The assumption is that parenting stress undermines sensitivity, but the inverse is also possible. Insensitive mothers might be confronted by related infant behaviors that then undermine their sense of parental competence, a key element of the construct of parenting stress. Further, effects aggregated here were unadjusted but the association between parenting stress and sensitivity is likely moderated by factors such as social support and partner relations (Pereira et al., 2012). As stress can spill-over within the marital or partner unit (Almeida, Wethington, & Chandler, 1999), parenting stress perceived by the infant’s father or co-parent might moderate effects within the triad. However, there was no paternal-report data on parenting stress available for aggregation. Research should be replicated in longitudinal study designs examining these possible mediating and moderating pathways.

**Quality Assessment of Included Studies**

As we highlighted in Table 3, studies utilized the MBQS with varying degrees of methodological rigor. A key quality indicator is transparent reporting of inter-rater reliability (IRR) and its calculation (i.e., how many cases were double-coded to generate an intra-class correlation and whether this was done by an experienced researcher). Due to high costs associated with lengthy observational research in this area, many labs are operated partly or largely by research students. Student-coded data potentially lack accuracy compared to data coded by experienced professionals. While IRR is an important indicator of reliability in these data, though, there is also likely to be undetectable variability across individuals in their capacity to detect nuances in sensitive caregiving.

There was also variability in measurement quality of stress indicators across studies. It is particularly important that measures of internalizing symptoms and parenting stress are drawn from reliable and valid instruments. However, even the most optimal study designs might not be equipped to capture underlying nuances in associations with sensitivity. In the assessment of internalizing symptomatology, for example, clinical diagnoses are considered more optimal than self-report data but might not always be sensitive to subclinical symptomatology that is potentially still relevant to maternal sensitivity.

In general, authors acknowledged sample limitations, but in some instances methodological bias was not reported. Measurement temporality of contextual factors was not reported for 10 of 30 aggregated effects. Many studies report sociodemographic data as a sample descriptor or confounder in analyses rather than a focal construct, and do not report on timing of measurement, particularly when embedded within a larger longitudinal study design.

We performed post-hoc sensitivity analyses where possible in order to determine whether various study or sample conditions could explain heterogeneity in findings. Sensitivity analyses demonstrated only negligible differences in aggregate effect sizes, and larger studies often drove our findings. Given the costs associated with research in this area, it remains important for findings from all (even small) samples to be published as we demonstrate that they can contribute to overall emerging associations.

**Strengths and Limitations of This Study**

Strengths of this review include the use of a standardized systematic search protocol following the PRISMA statement (Moher et al., 2009) and, where possible, the use of meta-analysis to generate pooled effect sizes. A limitation is that the systematic search was restricted to peer-reviewed studies. While we recognize that this may encourage oversampling of significant findings, contributing to the file drawer problem (Rosenthal, 1979), examining the grey literature retrieved from Google Scholar indicated that most unpublished studies came from the same labs as the (included) published studies. The necessary exclusion from meta-analyses due to overlapping samples would have disqualified much of this grey literature.

Effects included for meta-analysis were unadjusted, with the exception of one reported by Logsdon et al. (2015) indicating unipolar depressive symptoms, adjusted for race and marital status. However, this effect was similar in magnitude to most other unadjusted, effects of internalizing symptoms. Furthermore, we examined increases in levels of each contextual factor individually rather than assessing compounding levels of stress. In fact, there is probable interplay between these factors: many of the effects reported here are likely to include unmeasured shared variance with factors that are theoretically collinear. The aim of the current review was to grow an evidence base around the strength of aggregated bivariate stress effects, and available data were insufficient for testing any possible mediating or moderating effects of other theoretically relevant factors such as partner or social support.

Restricting the scope of this review to one operationalization of parental caregiving behavior excludes other domains such as mind-mindedness, reflective functioning, mutually responsive orientation and scaffolding. Future research should explore how these domains might differentially be affected by contextual stress. We also restricted our enquiry to a single cohesive operationalization of sensitivity coded by the Maternal Behavior Q-Sort. We saw the MBQS as particularly suitable for investigating contextual stress effects as it was specifically designed to account for caregiver capacity to attend to infant needs in a context of competing demands; it also demands highly standardized training requirements and demonstrates cross-cultural validity. However, this excludes an examination of stress associations with alternative instruments such as the Ainsworth Maternal Sensitivity Scales, the CARE-Index, the HOME Inventory and the Emotional Availability Scales, among others. Future research should expand on current findings by detecting whether contextual stress effects emerge in a broader literature consisting of diverse representations of sensitivity.

**Limitations in the Literature**

Overall, studies of maternal sensitivity lack focal empirical attention to indicators of contextual stress. More than half of retrieved studies that used the MBQS were ineligible for inclusion because they did not report a coefficient indicating an association between sensitivity and contextual stress factors. Many associations were unsuitable for meta-analysis of pooled effects, mostly because different studies drew from several common samples to answer various research questions related to maternal sensitivity. Much of the MBQS literature has emerged from three Canadian labs (Western Ontario; Montreal; and Quebec City). Because of this, there was no way to accurately assess the nature of sample overlap. Thus, we made the cautious decision to exclude additional, equivalent stress associations drawn from different studies that reported on common samples. It will be important for future studies to examine relationships in a diversity of cross-cultural samples.

Indicators of stress aggregated here were all measured in the perinatal period, with no preconception measures emerging as eligible for review. This limits our understanding of possible effects of preconception variables not examined here. For example, prolonged adversity across periods of development or adversity at particular time points during development might exert differential effects on later sensitivity. Alternatively, the effect of perinatal variables might be moderated by individual adolescent and preconception characteristics such as high levels of neuroticism or negative reactivity. Due to limited available longitudinal research, we could not assess any possible non-linear associations (for example, the relevance of a life history of adversity in the context of a supportive family environment). The experience of supportive environments where stress or adversity is well-managed might equip individuals with the emotion regulation capacity needed to maintain infant-oriented maternal sensitivity in spite of concurrent stress in the caregiving ecology.

Use of the MBQS has largely been restricted to samples of mothers. Further research is needed to clarify whether any gendered differences exist in parental sensitivity, particularly in an emerging sociocultural climate in which fathers and same-sex partners are increasingly becoming primary caregivers.

Maternal sensitivity qualitatively changes across time from early to mid-infancy and beyond, reflecting a child's shifting developmental needs (Feeney & Woodhouse, 2016; Solomon & George, 1996). As infants grow older, sensitive caregiving increasingly comprises support for learning and exploration (Feeney & Woodhouse, 2016; Wade et al., 2015). We did not restrict inclusion of effects based on infant age in this study, but it is important to bear in mind that contextual stress factors might be differentially relevant to dyadic functioning at different developmental time points.

**Conclusions and Recommendations**

We report important aggregated findings indicating that multiple conceptualisations of contextual stress demonstrate similar, negative associations with maternal sensitivity.

Clear relationships were found in each of our expected directions. The proposed mechanism is that contextual stress indicators examined here are commonly linked to maternal sensitivity through their effects on the ability to maintain motivational focus on child co-regulation in a context of competing demands or depleted psychological resources. The prototype sensitive parent can balance self-focused needs alongside the child's needs, but evidence suggests this may be an increasingly precarious equilibrium as ecological stress increases. We recommend that more attention is paid to ecological factors in maternal sensitivity research, and suggest as a matter of protocol that studies ensure transparency in the statistical reporting of sociodemographic and other contextual stress data. If contextual stressors are consistently found to undermine sensitive caregiving, then parents experiencing stress or adversity should be supported to develop strategies to maintain and act upon adaptive internal working models of caregiving sensitivity even when the capacity to meet their own needs is compromised.

This review supports the logical proposition that contextual stress is an important component of overall variation in maternal sensitivity. Aggregate effects of the stress ecology are related to only a portion of the variance in sensitivity (up to 10% for maternal age), but this is a key contribution considering that seminal meta-analytic evidence links 12% of the variance in maternal sensitivity to the theoretically pivotal contribution of adult attachment state of mind (van IJzendoorn, 1995). At this stage, it is of course unclear how much shared variance exists between these contributions. Future studies should also examine interrelations of contextual, representational, and biological/hormonal contributions to caregiving sensitivity.

We demonstrate smaller yet significant aggregated contributions of parenting stress and internalizing symptoms. Aggregated effects for these indicators might be conservative estimates given that when longer MBQS coding periods were observed, correlations with maternal sensitivity were stronger.

It is not surprising that while the settings of stress or adversity identified here exert significant effects on sensitivity, they do not completely undermine the caregiving system. The prototype optimal caregiver can withstand stressful intrusions and deftly attend to the needs of the infant, however, there must exist a threshold at which cracks in sensitivity become apparent. Further research should examine where vulnerability to ecological stress emerges and at what point it impedes a caregiver's ability to remain nondefensively open to the infant's emotion regulation needs. Research will be necessarily incremental in this area. This review highlights the need for further investigation at the foundational level whereby direct associations, along with the timing of stress demands and contingencies of sample characteristics are better understood. Then, examination of the mechanisms that underlie these associations should be explored, as well as possible moderators of stress factors and combined contributions of multiple characteristics of the family caregiving ecology. Pathways to caregiving sensitivity are expected to be complex and may become apparent using longitudinal methods that allow for examination of possible preconception moderators of perinatal sensitivity. A growing evidence base is needed to clarify the specific combinations and thresholds of contextual risk that will be in most need of support for optimizing sensitivity.

*This research did not receive any specific grant from funding agencies in the public, commercial, or not-for-profit sectors.*

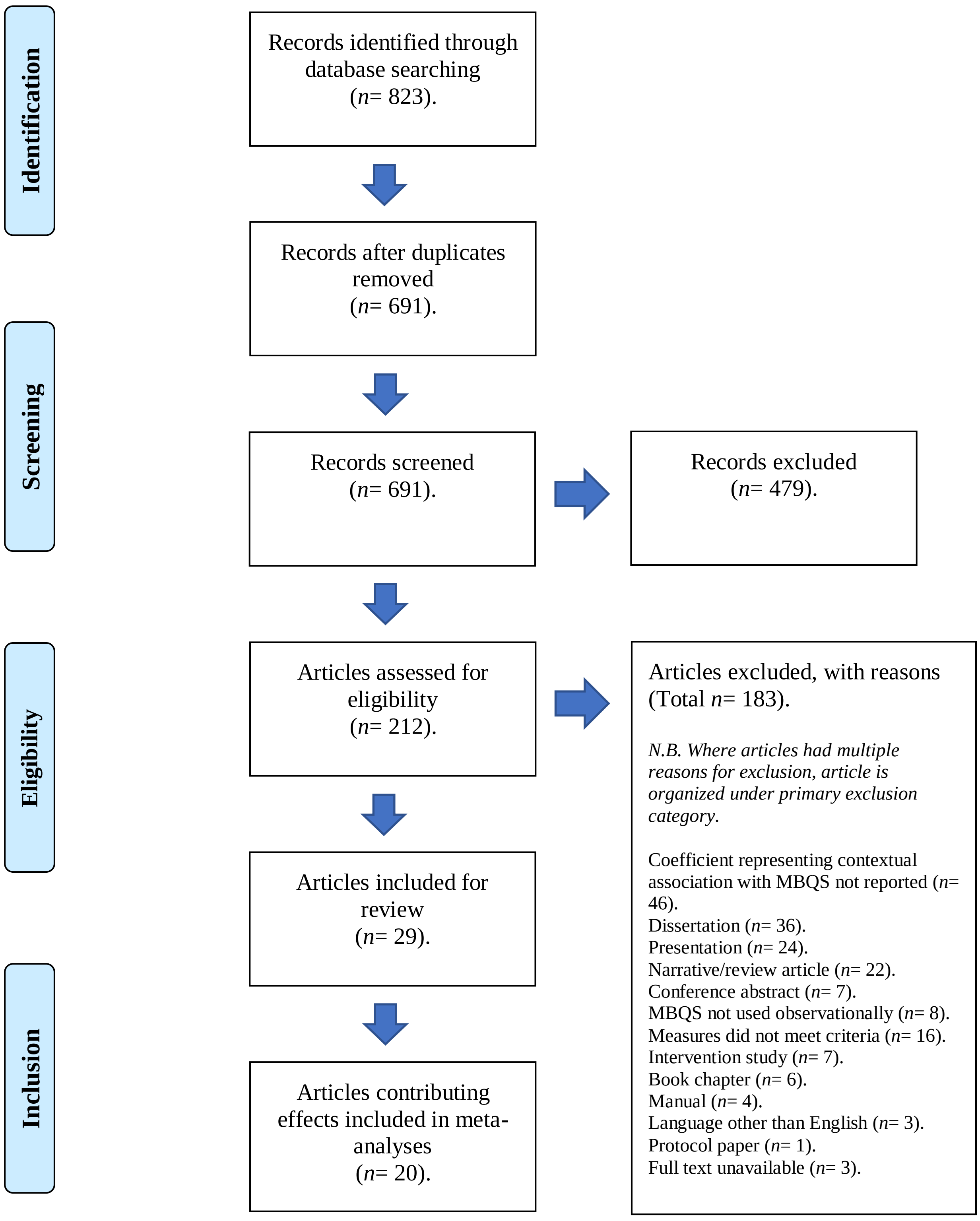

Fig. 1. Selection process of articles for inclusion in review and meta-analyses.

**In-text Tables**

Table 1
Recruitment and sample characteristics of included articles (*N*=29).

| Study | Recruitment | Participation criteria | *N*(*n*) | Maternal age range (years) | Child gender | Child age at MBQS observation (months) | Annual family income** mean (range) | Maternal education | Birth order/ siblings |
|---|---|---|---|---|---|---|---|---|---|
| Bailey et al. (2016) * | Postpartum from Southwestern Ontario hospitals. Longitudinal study sub-sample | Maternal age ≥20 years; infants born full term with no known medical complications | 66 | 20.2-44.85 (M=30.19, SD=4.93) | 31 f<br>35 m | 10 | M=$50,000 - $59,999 (Range=NR) | M=14.62 years (SD = 1.68) | No siblings |
| Bernier et al. (2010) * | From birth lists provided by Ministry of Health and Social Services | Full-term pregnancy; absence of disability in infant | 116 | 19-44 (M= 30.5) | 64 f<br>52 m | 12 | M=$70,000 (<$20,000 - >$100,000) | M=15 years | 59 children were firstborn; 43 had older sibling; 12 had two; 1 had three; 1 had four |
| Bernier & Matte-Gagne (2011) | From birth lists provided by Ministry of Health and Social Services | Full-term pregnancy; absence of physical/mental disability/developmental delay | 59 | 24-45 (M=31.9) | 30 f<br>29 m | 12 | M=$70,000 (<$20,000 - >$100,000) | 78%) had college degree | 30 of 59 children were first-born |
| Bernier et al. (2014) | From birth lists provided by Ministry of Health and Social Services | Full-term pregnancy; absence of disability in infant | 63 | 20-44 (M=31.8) | 27 f<br>36 m | M=12.5 (SD=0.9) | M=NR (<$20,000 - >$100,000) | M=15.2 years | NR |

| Study | Recruitment | Participation criteria | *N*(*n*) | Maternal age range (years) | Child gender | Child age at MBQS observation (months) | Annual family income** mean (range) | Maternal education | Birth order/ siblings |
|---|---|---|---|---|---|---|---|---|---|
| Bernier, McMahon & Perrier (2017) * | From birth lists provided by Ministry of Health and Social Services | Full-term pregnancy; absence of disability in infant | 204 | 20-45 (M=31.5) | NR | 12 | M=$60,000-$79,000 (Range= <$20,000- >$100,000) | M=15.6 years (Range= 8-18) | NR |
| Bigelow et al. (2010) * | Participants of randomized control study (*N*=34) of effects of skin to skin contact from birth versus incubator care. Infants born in public hospital, Cape Town | Infants born with low birth weight (1385-2199g) Subsample of 12: dyads still living together; known address; infant <12 month at follow-up | 12 | 16-30 (M=23, SD=5.6) | 5 f 7 m | M = 8 (SD = 3.3) (Range 3.1 to 11.9) | M=NR (Range=NR) | M=9.8 years (SD = 1.5) | NR |
| Bordeleau, Bernier & Carrier (2012) | From birth lists provided by Ministry of Health and Social Services | Full-term pregnancy; absence of any known physical or mental disability in infant | 70 | 20-45 (M=31.3) | 60% f 40% m | M = 12.8 | M=categorical score 4.34 (SD=1.5) (4=$60-79K; 5=$80-99K) | 65.2% had college degree | NR |
| Bouvette-Turcot, Bernier & Leblanc (2017) * | From birth lists provided by Ministry of Health and Social Services | Full-term pregnancy; absence of disability or severe delay in infant | 106 | 22-45 (M=31.7) | 55 f 51 m | 12 | M=$60,000-$79,000 (Range=NR) | M=16.1 (Range= 10-18) | 19% of children had no siblings |

| Study | Recruitment | Participation criteria | *N*(*n*) | Maternal age range (years) | Child gender | Child age at MBQS observation (months) | Annual family income** mean (range) | Maternal education | Birth order/ siblings |
|---|---|---|---|---|---|---|---|---|---|
| Brooker & Poulin-Dubois (2013) * | From birth lists provided by a governmental health agency | Toddler born full term with no signs of visual/auditory deficits, English or French as primary language in home | 42 | Range NR (M=34.78, SD=4.64) ('primary caregiver' aggregate includes fathers) | 19 f<br>23 m | M = 24.31 (SD = 0.61) | M=$75,000-$100,000 (Range <$22,000->$150,000) | Bachelor's level of education on average ('primary caregiver' aggregate includes fathers) | NR |
| Dayton, Huth-Bocks & Busuito (2016) | Through fliers posted at public locations and agencies for low-income families | Subsample of larger longitudinal study examining psychosocial and environmental risk | 114<br>99 | 18-42 (M=26, SD=5.7) | NR | i) 12<br>ii) 24 | Median=$1500/ month | 20 high school diploma or less; 44 some college/trade school; 36 college degree | 47.5% primiparous |
| Deschenes et al. (2014) | From birth lists provided by Ministry of Health and Social Services | Full-term pregnancy; absence of known developmental delays | 74 | 20-45 (M=31.6, SD=4.95) | 46 f<br>28 m | M = 12.5 (SD = 1.11) | M=categorical score 4.4 (SD=1.5) (4=$60-79K; 5=$80-99K) | M=15.60 years (SD = 0.25) | 33 infants were first-born; 24 had one sibling; 12 had two; 4 had three; 1 had four |

| Study | Recruitment | Participation criteria | *N*(*n*) | Maternal age range (years) | Child gender | Child age at MBQS observation (months) | Annual family income** mean (range) | Maternal education | Birth order/ siblings |
|---|---|---|---|---|---|---|---|---|---|
| Khoury et al. (2016) * | From community centres, mother-infant activity centres, and consumer baby shows across urban and suburban area | Maternal age ≥18 years at childbirth; no known hormonal disorders; fluent in English; infants with no major developmental disorder and born of pregnancy >32 weeks | 297 | 21.36-46.52 (M=33.44, SD=4.51) | 152 f<br>145 m | 16 | Median $70,000 - $140,000 | 48% had university degree | NR |
| Lemelin, Tarabulsy & Provost (2006) | Via maternity ward nurses of major birthing hospital of Quebec and visiting nurses of local community health centres | For both groups: infants born without physical or congenital anomalies | 89 | Adults: >20 (M=29.74, SD=4.65)<br>Adolescents: ≤19 (M=18.39, SD=1.58) | Adult dyads:<br>12 f<br>15 m<br>Adolescent dyads:<br>31 f<br>31 m | i) 15<br>ii) 18 | Adults: M≥$30K (for 81% sample) (Range=NR)<br>Adolescents: M≤$30K (for 85% sample) (Range=NR) | Adults: M=15.48 years (SD = 2.39)<br>Adolescents: M=9.98 years (SD = 1.58) | Adult dyads: 13 infants were firstborn<br>Adolescent dyads: 53 infants were firstborn |
| Logsdon et al. (2015) * | Self-referral, physician and community health centre referral, and/or advertising, ≤20 weeks' gestation | i) Diagnosis of bipolar depression<br>ii) Diagnosis of unipolar depression<br>iii) Control (no diagnosis) | 90 | 18-44 | NR | 12 | M=NR (Range=NR) | NR | NR |

| Study | Recruitment | Participation criteria | *N*(*n*) | Maternal age range (years) | Child gender | Child age at MBQS observation (months) | Annual family income** mean (range) | Maternal education | Birth order/ siblings |
|---|---|---|---|---|---|---|---|---|---|
| Moran et al. (1992) * | Participation in intervention & support program for increasing maternal sensitivity at Children's Psychiatric Research Institute in London, Ontario | Infant able to locomote independently; participation in program for at least 6 months | 19 | 19-35 (M=30) | NR | M = 20 | M=NR (Range=NR) | M=13 years | NR |
| Pederson et al. (1990) * | From volunteer subject pool maintained by university child development study group and newspaper birth announcements | NR | 40 | 22-39 (M=29.9) | NR | 12 | M=$35,000 (Range=NR) | M=14.9 years | Families included 1 to 4 children (M=2) |

| Study | Recruitment | Participation criteria | *N*(*n*) | Maternal age range (years) | Child gender | Child age at MBQS observation (months) | Annual family income** mean (range) | Maternal education | Birth order/ siblings |
|---|---|---|---|---|---|---|---|---|---|
| Pereira et al. (2012) * | From community centres, mother-infant activity centres, and consumer baby shows across urban and suburban area | Mothers ≥18 years at childbirth, no known psychiatric disorder, sufficient English to complete questionnaires | 291 | 33.38 (SD=4.35) | 49.1% f 50.9% m | 16 | Median= $114,000-149,999 (Range=NR) | NR | 57.2% of infants had no siblings; 17.1% had 2 or 3; 0.9% had 3 or 4 siblings |
| Posada et al. (2004) * | Through health, housing and education provider | NR | 30 | 23-39 (M=31.4) | 16 f 14 m | M = 13.1 (range 8–24) | M=NR (Range=NR) | 1 incomplete high school; 8 high school degree; 7 technical degree; 14 uni degree | Infants had one sibling on average (range 0-4) |
| Rifkin-Graboi et al. (2015) * | Larger (GUSTO) cohort of pregnant women attending first trimester clinic at National University Hospital and KK Women's and Children's Hospital, Singapore | Infants: Apgar score of ≥9, gestational age ≥37 and <42 weeks, birth weight ≥2500 and <4000g, singleton birth and born to mothers with no pregnancy complications | 271<br>261<br>227 | NR | NR | 6 | M=NR (Range=NR) | NR | NR |

| Study | Recruitment | Participation criteria | *N*(*n*) | Maternal age range (years) | Child gender | Child age at MBQS observation (months) | Annual family income** mean (range) | Maternal education | Birth order/ siblings |
|---|---|---|---|---|---|---|---|---|---|
| Rochette & Bernier (2014) | From birth lists provided by Ministry of Health and Social Services | NR | 114 | 20-45 (M=31.41, SD=4.99) | 68 f<br>46 m | M = 12.58 | M=$74,000 (close to M family income for Canada) (Range=NR) | M=15.6 years (SD = 2.36) | NR |
| Tarabulsy et al. (2005) * | Via maternity ward nurses from major birthing hospital and visiting nurses from local community health centres | Maternal age ≤19 years | 64 | M=17.4, SD=1.5 | 33 f<br>31 m | i) 6<br>ii) 10 | M=$0-15K (assessed at 2 time pts, when infants were 6 and 18 months old) (Range=NR) | M=10.1 years (SD = 1.3) | NR |
| Tarabulsy et al. (2008) * | Via maternity ward nurses of 2 major birthing hospitals and with the help of public health nurses of local community health centres | For both groups: infants born weighing more than 2500g without physical or congenital abnormalities | 127 | Adults: M=28.8 (SD=4.66)<br>Adolescents: M=18.07 (SD=1.25) | Adult dyads:<br>18 f<br>26 m<br>Adolescent dyads:<br>41 f<br>42 m | i) 6<br>ii) 10 | Adults: M=$30-45K (Range=NR)<br>Adolescents: M=$0-15K (Range=NR) | Adults: M=14.36 years (SD = 3.54)<br>Adolescents: M=9.92 years (SD = 1.72) | NR |

| Study | Recruitment | Participation criteria | *N*(*n*) | Maternal age range (years) | Child gender | Child age at MBQS observation (months) | Annual family income** mean (range) | Maternal education | Birth order/ siblings |
|---|---|---|---|---|---|---|---|---|---|
| Tarabulsy et al. (1997) | Mothers recruited shortly after infant birth at a hospital in London, Ontario | Preterm infants selected from neonatal ICU if birth weight <2000g; gestational age <37 weeks; and absence of gross neurological or physical anomalies at birth | 79 | Preterm mothers: M=29 (SD=4.9) Full-term mothers: M=30 (SD=4.9) | NR | 8 | Preterm families: M= $50K (Range=NR) Fullterm families: M= $50K (Range=NR) | Preterm mothers: M=14 years (SD = 2.7) Fullterm mothers: M=15 years (SD = 2.7) | NR |
| Wade et al. (2015) * | Through Healthy Babies Healthy Children program (contacts parents of all newborns within days of each birth). Births between Feb 2006 and Feb 2008 in Toronto & Hamilton | For participation in IKFP subsample (Kids, Families and Places intensive sample): English-speaking mother; newborn weight >1500g; two or more children <4 years old | 379 | NR | 247 f 254 m | 2 | M=$30,000-39,999*** (census population mean = $30,504.16) (Range=NR) | 53% had bachelor's degree | Families included a newborn (target child at recruitment point) plus two or more children <4 years old |

| Study | Recruitment | Participation criteria | *N*(*n*) | Maternal age range (years) | Child gender | Child age at MBQS observation (months) | Annual family income** mean (range) | Maternal education | Birth order/ siblings |
|---|---|---|---|---|---|---|---|---|---|
| Wen et al. (2017) | Eligible subsample of Growing Up in Singapore Towards Healthy Outcomes (GUSTO) longitudinal birth cohort study | Healthy term-born infants with gestational age ≥37 weeks and birth weight ≥2.5kg; 5-minute neonatal APGAR score ≥9 | 111 | NR | NR | 6 | NR | NR | NR |
| Whipple, Bernier & Mageau (2011) | Random recruitment through birth lists provided by Quebec Ministry of Health and Social Services | Full-term pregnancy; absence of severe developmental delays | 71 | 20-45 (M=31, SD=4.7) | 34 f<br>37 m | 12 | M= categorical score 4.04 (SD=1.6) (4 = $60,000 - $79,000; 5 = $80,000 - $99,000) | M=15 years (SD = 2.5) | 36 infants were firstborn; 35 infants had older siblings |
| Whipple, Bernier & Mageau (2011) (2) | Random recruitment through birth lists provided by Quebec Ministry of Health and Social Services | Full-term pregnancy; absence of severe developmental delays | 71 | M=30.8, SD=4.5 | 37 f<br>34 m | 12 | M=NR (< $20,000 to > $100,000) | M=15 years (SD = 2.4) | NR |

| Study | Recruitment | Participation criteria | *N*(*n*) | Maternal age range (years) | Child gender | Child age at MBQS observation (months) | Annual family income** mean (range) | Maternal education | Birth order/ siblings |
|---|---|---|---|---|---|---|---|---|---|
| Xing et al. (2017) * | Through information posted on websites and in large Beijing communities | First-born child; full-term pregnancy; infants with no physical or mental disability; infant between 12-24 months of age; mother employed outside the home; child cared for by grandmothers for >10 hrs/week | 71 | NR | 33 f 37 m | M=17.6 (SD=3.73) | 34.7% ≤6000 Yuan; 37.5% 6,000 – 10,000 Yuan; 27.8% > 10,000 Yuan | 61.1% university educated; 25% graduate education; 13.9% high school education | All infants first-born |
| Zreik, Oppenheim & Sagi-Schwartz (2017) * | By telephone based on birth records provided by Ministry of the Interior | Dyads living in either of two large Arab cities in northern Israel; healthy infants born ≥36 weeks' gestation and weighing ≥2500g at birth | 76 | Range NR; M=29.91 (SD=4.85) | 44 f 41 m | 12-18 months (M=13.97; SD= 1.87) | NR | M=15.05 years (SD= 3.01) | 31 (36.5%) of children were first-born; families had average 2.04 children |

*Note*. NR = Not Reported.
*Included for meta-analysis
**All income reported in Canadian dollars unless otherwise specified
***Mean personal income

Table 2
Overall study predictors and outcomes, and findings specific to contextual stress association (*N*=29 studies).

| Study | Predictors/ moderators/ covariates | Outcome/s | Contextual stress indicator/s | Reported findings specific to contextual association with maternal sensitivity (MBQS) |
|---|---|---|---|---|
| Bailey et al. (2016) * | MBQS; Personality Assessment Inventory | Infant attachment security | Maternal depressive symptoms | Sensitivity sig. associated with depressive symptoms at 10 months (*r*=.26, *p*<.05) Sensitivity not sig. associated with depressive symptoms at 4 months (*r*=.04, *ns*) |

| Study | Predictors/ moderators/ covariates | Outcome/s | Contextual stress indicator/s | Reported findings specific to contextual association with maternal sensitivity (MBQS) |
|---|---|---|---|---|
| Bernier et al. (2010) * | Pregnancy, delivery and neonatal factors; maternal psychological distress | MBQS | Maternal age; maternal education; paternal education; family income; maternal psychological distress | Sensitivity sig. associated with maternal education (*r*=.29, *p*<.01), paternal education (*r*=.29, *p*<.01), family income (*r*=.37, *p*<.001) and psychological distress (*r*=-.21, *p*<.05) Sensitivity not sig. associated with maternal age (*r*<.12, *ns*) |
| Bernier & Matte-Gagne (2011) | Adult attachment (AAI); romantic attachment styles; maternal education | MBQS; infant attachment security; marital satisfaction | Maternal education | Sensitivity sig. associated with maternal education (*r*=.21, *p*<.05) |
| Bernier et al. (2014) | MBQS; sleep | Sleep; infant attachment security; infant theory of mind; infant executive functioning | SES | Sensitivity sig. associated with SES (*r*=.30, *p*<.05) |
| Bernier, McMahon & Perrier (2017) * | Maternal mind-mindedness; MBQS; child cognitive ability; family SES | Children's school readiness | SES | Sensitivity sig. associated with SES (*r*=.28, *p*<.001) |
| Bigelow et al. (2010) * | Mother-infant skin-to-skin contact | MBQS; NCATS | Maternal age; cohabiting with father; maternal education; maternal employment; number of people living in house | Sensitivity not sig. associated with maternal age (*r*=.37, *ns*), education (r=.45, *ns*), cohabitation (*r*=-.02, *ns*), maternal employment (*r*=.38, *ns*) or the number of people living together in the house (*r*=.29, *ns*) |
| Bordeleau, Bernier & Carrier (2012) | MBQS; maternal mind-mindedness; maternal autonomy support; paternal MRO | Children's sleep at preschool age | Family income; maternal education | Sensitivity sig. associated with family income and maternal education (*r* range= .22-.34, *p*<.05) |

| Study | Predictors/ moderators/ covariates | Outcome/s | Contextual stress indicator/s | Reported findings specific to contextual association with maternal sensitivity (MBQS) |
|---|---|---|---|---|
| Bouvette-Turcot, Bernier & Leblanc (2017) * | Maternal psychosocial maladjustment; MBQS | Child internalizing symptoms | Maternal age; maternal parenting stress; maternal education; paternal age; paternal education; family income; maternal relationship satisfaction (Dyadic Adjustment Scale); maternal psychiatric symptoms; maternal psychosocial maladjustment | Sensitivity sig. associated with maternal age (*r*=.20, *p*<.01), maternal education (*r*=.22, *p*<.01), paternal education (*r*=.20, *p*<.01), family income (*r*=.24, *p*<.01)<br>Sensitivity not sig. associated with maternal parenting stress (*r*=-.03, *ns)*, paternal age (*r*=.07, *ns*), maternal Dyadic Adjustment Scale total (relationship satisfaction) (*r*=-.01, *ns*), maternal Psychiatric Symptoms Index total (*r*=.01, *ns*), or maternal psychosocial maladjustment composite (r=-.06, *ns*) |
| Brooker & Poulin-Dubois (2013) * | MBQS | Toddlers' word learning, imitation and instrumental helping | Income; parental education | Sensitivity not sig. associated with income (*r*=.24, *ns*) or parental education (*r*=.25, *ns*) |
| Dayton, Huth-Bocks & Busuito (2016) | Maternal interpersonal aggression exposure | Perceptions of infant emotion; MBQS | Romantic relationship quality (Marital Relationship Scale; MRS); intimate partner aggression (IPA) | Sensitivity sig. associated with MRS: Ambivalence subscale (12 months) (*r*=-.23, *p*<.05), MRS: Love subscale (24 months) (*r*=.39, *p*<.01), MRS: Ambivalence subscale (24 months) (*r*=-.30, *p*<.01), IPA during pregnancy (24 months) (*r*=-.28, *p*<.01)<br>Sensitivity not sig. associated with MRS: Love subscale (12 months) (*r*=.16, *ns*), MRS: Conflict subscale (12 months) (*r*=-.05, *ns*), MRS: Conflict subscale (24 months) (*r*=-.02, *ns*), IPA during pregnancy (12 months) (*r*=-.40, *ns*), IPA year before pregnancy (12 months) (*r*=.00, *ns*), IPA 1st year postpartum (12 months) (*r*=.05, *ns*), IPA year before pregnancy (24 months) (*r*=-.12, *ns*), IPA 1st year postpartum (24 months) (*r*=-.06, *ns)* |

| Study | Predictors/ moderators/ covariates | Outcome/s | Contextual stress indicator/s | Reported findings specific to contextual association with maternal sensitivity (MBQS) |
|---|---|---|---|---|
| Deschenes et al. (2014) | Family SES; child sex; siblings; MBQS; MRO | Similarity in MBQS (mothers) and MRO (fathers) | Family SES | Sensitivity sig. associated with family SES (*r*=.37, *p*<.001) |
| Khoury et al. (2016) * | Maternal depressive symptoms; maternal cortisol levels | Infant cortisol levels | Maternal depression; ethnicity | Sensitivity not sig. associated with maternal depression (*r*=-.01, *ns*) or ethnicity (*r*=-.02, *ns*) |
| Lemelin, Tarabulsy & Provost (2006) | MBQS; infant temperament; psychosocial risk | MBQS; preschool cognitive development | Psychosocial risk | Sensitivity sig. associated with psychosocial risk (*r*=-.32, *p*<.01) |
| Logsdon et al. (2015) * | Maternal depression | MBQS; AMSS; Dyadic Mini Code; Child-Caregiver Mutual Regulation Scale | Maternal depression | Sensitivity not sig. associated with unipolar depression diagnosis (*r*=-.12, *ns*) (converted effect); sensitivity scores lower in women with bipolar than unipolar depression or control group |
| Moran et al. (1992) * | MBQS; HOME; Bromwich PBP; Ainsworth scales; developmental delay diagnosis | Infant attachment security (AQS); PSI | Maternal age; maternal education; family SES | Sensitivity not sig. associated with SES (*r*=.38, *ns*), maternal education (*r*=.24, *ns*) or maternal age (*r*=.23, *ns*) |
| Pederson et al. (1990) * | MBQS; Ainsworth scales; PSI | Infant attachment security (AQS) | Maternal education; maternal parenting stress; maternal age; income; | Sensitivity sig. associated with maternal education (*r*=.29, *p*<.05) and PSI Child Domain (*r*=-.36, *p*<.05)<br>Sensitivity not sig. associated with PSI Parent Domain (*r*=-.10), maternal age, income, or SES (*r* NR) |
| Pereira et al. (2012) * | Childhood Trauma Questionnaire; PSI-SF | MBQS | Maternal parenting stress | Sensitivity sig. associated with maternal parenting stress (parent domain) (*r*=-.13, *p*<.05); (child domain) *r*=-.16; (interaction domain) *r*=-.13; PSI (total) *r*=-.16 |

| Study | Predictors/ moderators/ covariates | Outcome/s | Contextual stress indicator/s | Reported findings specific to contextual association with maternal sensitivity (MBQS) |
|---|---|---|---|---|
| Posada et al. (2004) * | Open-ended observations of maternal caregiving behavior; MBQS | AQS | Maternal education | Sensitivity not sig. associated with maternal education (*r*=.03, *ns*) |
| Rifkin-Graboi et al. (2015) * | MBQS | Infant brain structure and function | Maternal education; household income; maternal anxiety | Sensitivity sig. associated with household income (*r*=.21, *p*<.01) Sensitivity marginally associated with maternal education (*r*=.11, *p*=0.058); sensitivity at infant age 6 months marginally associated with maternal state anxiety at infant age 3 months (*r*=-.10, *p*=0.116) |
| Rochette & Bernier (2014) | MBQS; family SES | Child executive functioning | SES | (Disaggregated MBQS domains): response to distress sig. associated with SES (*r*=.20, *p*<.05); sensitivity/ responsiveness sig. associated with SES (*r*=.23, *p*<.05); physical proximity sig. associated with SES (*r*=.20, *p*<.05) Response to positive signals marginally associated with SES (*r*=.17, *p*<.10) Positive affect sharing not sig. associated with SES (*r*=.05, *ns*); hostility/rejection not sig. associated with SES (*r*=-.03, *ns*) |
| Tarabulsy et al. (1997) | MBQS; AQS; SSP; PSI; Infant Characteristics Questionnaire | (Group comparisons) | Parenting stress (child domain) | Sensitivity not sig. associated with parenting stress (child domain) (*r*=-.04, *ns*) |
| Tarabulsy et al. (2005) * | AAI; MBQS (mediator); ecological variables | AQS; MBQS (mediator) | Maternal education; maternal depression; maternal satisfaction with social support | Sensitivity sig. associated with maternal education (*r*=.39, *p*<.001) Sensitivity not sig. associated with maternal depression (*r*=.00, *ns*), maternal satisfaction with social support from infant's father (*r*=.09, *ns*), from grandmother (*r*=.10, *ns*) or general support (*r*=.09, *ns*) |

| Study | Predictors/ moderators/ covariates | Outcome/s | Contextual stress indicator/s | Reported findings specific to contextual association with maternal sensitivity (MBQS) |
|---|---|---|---|---|
| Tarabulsy et al. (2008) * | MBQS; infant temperament; parental stress; psychosocial risk; AQS | AQS | Maternal age; parenting stress; life events | Sensitivity sig. associated with psychosocial risk (*r*=-.44, *p*<.001) Sensitivity marginally associated with life events (*r*=-.15, *p*<.10) Sensitivity not sig. associated with parenting stress (*r*=-.10, *ns*) |
| Wade et al. (2015) * | Cumulative risk index (maternal education, depression, history of abuse in childhood, lack of organization and safety in the home); MBQS | MBQS; Parent-Child Interaction System; CARP; Infant cognitive competency (various dimensions) | Cohabiting with father; maternal education; maternal depression; step-family status; cumulative risk index; organisation and safety in the home | Sensitivity sig. associated with maternal education (*r*=.22, *p*< .01), absence of infant's father from home (*r*=-.12, *p*<.05), maternal depression (*r*=-.16, *p*<.01), organisation and safety in the home (*r*=.15, *p*<.05), and a cumulative risk index (*r*=-.20, *p*<.01) Sensitivity not sig. associated with step-family status (*r*=-.06, *ns*) |
| Wen et al. (2017) | Maternal depression; MBQS | Infant frontal EEG asymmetry | Family income; maternal education | Sensitivity sig. associated with family income (*r*=.31, *p*<.01) and maternal education (*r*=.25, *p*<.01) |
| Whipple, Bernier & Mageau (2011) | AAI; SES | MBQS; maternal autonomy support | Maternal age; SES | Sensitivity sig. associated with SES (*r*=.29, *p*<.05) Sensitivity not sig. associated with maternal age (*r*=.02, *ns*) |
| Whipple, Bernier & Mageau (2011) (2) | MBQS; SES; autonomy support | AQS | SES | Sensitivity sig. associated with SES (*r*=.32, *p*<.01) |
| Xing et al. (2017) * | MBQS (maternal and grand-maternal) | Infant temperamental reactivity; infant behavior problems | Maternal education; income | Sensitivity sig. associated with maternal education (*r*=.40, *p*<.01) and income (*r*=.38, *p*<.01) |
| Zreik, Oppenheim & Sagi-Schwartz (2017) * | MBQS | Infant attachment | Maternal education | Sensitivity sig. associated with maternal education (*r*=.32, *p*<.01) |

*Note*. Table includes several contextual indicators that were not eligible for aggregation as they were single study effects (e.g., ethnicity; intimate partner aggression). MBQS=Maternal Behavior Q-Sort; SES=socioeconomic status; AAI=Adult Attachment Interview; AQS=Attachment Q-Sort; MRO=Mutually Responsive Orientation; CARP=Coding of Attachment-Related Parenting; PSI=Parenting Stress Index.
*Included for meta-analysis

Table 3
Summary of MBQS characteristics for included studies (*N*=29 studies).

| Study | MBQS version | Observation time; Environment | Intra-class correlation coefficient; N coders; N cases compared | Mean (SD); Range of scores |
|---|---|---|---|---|
| Bailey et al. (2016) * | 90 item & 25 item | 2 hours In-home | ICC = .82 Between 2 coders On 12 cases (20%) | M = .33 (.52); Range NR |
| Bernier et al. (2010) * | 90 item | 75-90 minutes In-home | ICC = .89 Coders not reported On 25 cases (21.6%) | NR; NR |
| Bernier & Matte-Gagne (2011) | 90 item | 90 minutes In-home | ICC = .87 Between 2 coders On 21 cases (36%) | M = .62 (.33); Range = -.82 - .87 |
| Bernier et al. (2014) | 90 item | 90 minutes In-home | ICC = .86 Between 2 coders On 13 cases (20.6%) | M = .66 (.28); Range = -.60 - .89 |
| Bernier, McMahon & Perrier (2017) * | 90 item | 10-minute free play plus a range of additional tasks In-home | ICC = .86 Between 2 coders On 20% of cases | NR; NR |
| Bigelow et al. (2010) * | 90 item | M = 37 minutes (SD = 10 minutes) In-home | ICC = .82 Between 2 coders No. cases not reported | NR; NR |
| Bordeleau, Bernier & Carrier (2012) | 90 item | 20 minutes' free play plus additional competing demands tasks etc. In-home | ICC = .87 Between 2 coders On 22% of cases | M = 0.62 (0.31); Range = -.33 - .87 |

| Study | MBQS version | Observation time; Environment | Intra-class correlation coefficient; N coders; N cases compared | Mean (SD); Range of scores |
|---|---|---|---|---|
| Bouvette-Turcot, Bernier & Leblanc (2017)<br>* | 90 item | 20 minutes' free play plus additional competing demands tasks etc.<br>In-home | ICC = .87<br>Between 2 coders<br>On 30% of cases | M = 0.64 (0.27); Range NR |
| Brooker & Poulin-Dubois (2013)<br>* | 25 item | 10 minutes<br>Lab | ICC = .91<br>Between 2 coders<br>On 20% of cases | M = 0.68 (0.25); Range = -0.15 to 0.87 |
| Dayton, Huth-Bocks & Busuito (2016) | 25 item | NR;<br>In-home | NR<br>(Group codes)<br>NR | 12 months: M = 0.41 (0.43); Range NR<br>24 months: M = 0.36 (0.49); Range NR |
| Deschenes et al. (2014) | 90 item | 75 minutes<br>In-home | NR<br>NR<br>NR | M = 0.68 (0.23); Range = -0.18 - 0.89 |
| Khoury et al. (2016)<br>* | 90 item | NR;<br>In-home | ICC = .88<br>Between 2 coders<br>Cases NR | NR; NR |
| Lemelin, Tarabulsy & Provost (2006) | 90 item | 2-3 hours<br>In-home | At 15 months:<br>ICC = .92<br>Between 2 coders<br>On 27 cases (30%)<br>At 18 months:<br>ICC = .84<br>Between 2 coders<br>On 11 cases (12%) | M = 0.34 (0.40); Range NR |
| Logsdon et al. (2015)<br>* | NR | 3 minutes<br>Lab | NR<br>NR<br>NR | NR; NR |

| Study | MBQS version | Observation time; Environment | Intra-class correlation coefficient; N coders; N cases compared | Mean (SD); Range of scores |
|---|---|---|---|---|
| Moran et al. (1992)<br>* | 90 item | 90 minutes<br>In-home | ICC = .97<br>Between 2 coders<br>On 13 cases (68%) | M = .13 (.54); Range NR |
| Pederson et al. (1990)<br>* | 90 item | 2 hours<br>In-home | ICC = .75<br>Between 2 coders<br>On 40 cases (100%) | M = .73 (.18); Range NR |
| Pereira et al. (2012)<br>* | 90 item | 2 hours<br>In-home | ICC = .88<br>Between 2 coders<br>Cases NR | M = .47 (.34); Range = -.69 - .90 |
| Posada et al. (2004)<br>* | 90 item | 2 hours<br>In-home | ICC = .85<br>Between 2 coders<br>Cases NR | M = .69 (.14); Range = .23 - .87 |
| Rifkin-Graboi et al (2015)<br>* | 25 item | 15 minutes<br>Lab | ICC = .94<br>Coders NR<br>On 70% cases | NR; NR |
| Rochette & Bernier (2014) | 90 item | 70-90 minutes<br>In-home | ICC = .84<br>Between 2 coders<br>On 30 cases (26%) | Response to positive signals: M = 7.25 (1.13); Range = 2.36-8.45<br>Response to distress: M = 7.18 (1.32); Range = 1.71-8.57<br>Positive affect sharing: M = 7.35 (1.23); Range = 1.43-8.86<br>Hostility/rejection: M = 2.92 (1.05); Range = 1.38-7.38<br>Sensitivity/responsiveness: M = 6.49 (0.84); Range = 3.15-7.48<br>Physical proximity: M = 6.84 (1.14); Range = 2.00-8.14 |
| Tarabulsy et al. (1997) | 90 item | 2-3 hours<br>In-home | NR | M = .29 (.46); Range NR |

| Study | MBQS version | Observation time; Environment | Intra-class correlation coefficient; N coders; N cases compared | Mean (SD); Range of scores |
|---|---|---|---|---|
| Tarabulsy et al. (2005) * | 90 item | Time NR In-home | ICC = .93 Coders NR On 20 cases (31%) | M = .33 (.32); Range = -.41 - .83 |
| Tarabulsy et al. (2008) * | 90 item | 2-3 hours In-home | 6 months (T1): ICC = .91 Between 2 coders On 17 cases (13%) 10 months (T2): ICC = .86 Between 2 coders On 29 cases (22%) | M = .40 (.34); Range NR |
| Wade et al. (2015) * | NR | 15 minutes In-home | ICC = .82 Between 6 coders Cases not reported | M = 0.28 (0.49); Range NR |
| Wen et al. (2017) | 25 item | 15 minutes Lab | ICC = .86 Between 2 coders On 64 cases | M = 0.26 (0.46); Range -0.76 – 0.90 |
| Whipple, Bernier & Mageau (2011) | 90 item | 90 minutes In-home | ICC = .89 Between 2 coders On 25 cases (36%) | NR; NR |
| Whipple, Bernier & Mageau (2011) (2) | 90 item | 90 minutes In-home | ICC = .85 Coders NR On 14 cases (20%) | M = .59 (.34); Range = -.60 - .86 |
| Xing et al. (2017) * | 72 item | 30-40 minutes In-home | ICC = .72 Between 2 coders Cases NR | NR; NR |

| Study | MBQS version | Observation time; Environment | Intra-class correlation coefficient; N coders; N cases compared | Mean (SD); Range of scores |
|---|---|---|---|---|
| Zreik, Oppenheim & Sagi-Schwartz (2017) * | 72 item | 30 minutes plus additional episodes<br>In-home | ICC = .80<br>Between 2 coders<br>On 21% of cases | M = 0.35 (0.57); Range NR |

*Note*. NR = Not Reported.
*Included for meta-analysis

Table 4
Measurement of contextual indicators included for meta-analysis.

| Indicator (*N* effects) | Definition | Measurement method and tool |
|---|---|---|
| Maternal education (9) | Total length of mother's education in years | Self-report: questionnaire (*N*=4)<br>NR (*N*=5) |
| Income (4) | Mean annual family income | Self-report: questionnaire (*N*=4) |
| Composite SES (2) | Aggregate indicator of family socioeconomic status | Self-report: questionnaire: standardized average of maternal education, paternal education and family income (*N*=1)<br>NR: SES index based on census data (Blishen et al., 1987) (*N*=1) |
| Maternal age (4) | Age of mother in years | NR (*N*=4) |
| Cohabitation (2) | Whether mother is living with the child's father | NR (*N*=2) |

| Indicator (*N* effects) | Definition | Measurement method and tool |
|---|---|---|
| Maternal internalizing symptoms (7) | Mother's experience of mental health problems including depressive symptoms, anxiety, irritability, cognitive disturbance, unipolar depression diagnosis | NR (*N*=2)<br>Self-report: Centre for Epidemiologic Studies Depression Scale (CES-D) 20-item (*N*=1)<br>Self-report: Beck Depression Inventory (BDI-II) 21-item (*N*=1)<br>Self-report: Depression Anxiety Stress Scales (DASS) depression scale scored as sum across 14 subscale items (*N*=1)<br>Self-report: Psychiatric Symptom Index (PSI) encompassing anxiety, depression, irritability and cognitive disturbance (*N*=1)<br>Clinician diagnosis: Structured Interview Guide for the Hamilton Scale for Depression with Atypical Depression Supplement (*N*=1) |
| Parenting stress (2) | Mother's experience of distress in the parenting role | Self-report: Parenting Stress Index (PSI) (long form) (*N*=1)<br>Self-report: Parenting Stress Index (PSI-SF) (short form) 36-item (*N*=1) |

*Note*. NR = Not Reported.

Table 5
Quality assessment of included studies (*N*=29).

| Study | Reporting of sample bias | Coder training reported | Type of coder | Competing demands | Reported IRR/ICC | Reported temporality | Instrument used to measure contextual indicator |
|---|---|---|---|---|---|---|---|
| Bailey et al. (2016) * | Relatively small sample size not overtly acknowledged.<br>Details of subsample extraction from broader longitudinal study not reported.<br>Comparatively low income not explained.<br>Acknowledged that preliminary findings require replication. | No | 'Researchers trained in home observations' | NR | Yes | Yes | Self-report depressive symptoms (DASS) |

| Study | Reporting of sample bias | Coder training reported | Type of coder | Competing demands | Reported IRR/ICC | Reported temporality | Instrument used to measure contextual indicator |
|---|---|---|---|---|---|---|---|
| Bernier et al. (2010) * | Did not specify whether recruitment from birth lists was random. Acknowledged lack of maternal mental health measure concurrent to other assessments (link to potentially deficient dyadic interactions at later assessment). | No | 'Extensively trained research assistants' | Yes | Yes | Yes | Self-report psychological distress (PSI) |
| Bernier & Matte-Gagne (2011) | Acknowledged relatively small sample. Acknowledged limitations associated with sample of mothers only. Specified that birth list recruitment was random. | Yes | 'Trained research assistants' | Yes | Yes | No | N/A |
| Bernier et al. (2014) | Acknowledged modest sample size. Acknowledged sample was well-educated and findings might not generalize to lower SES families, but does not report family income mean. Specified that birth list recruitment was random. | Yes | 'Trained graduate observers' | Yes | Yes | No | N/A |
| Bernier, McMahon & Perrier (2017) * | Specified that birth list recruitment was random. Specified details of attrition characteristics and missing data estimation. | No | 'Trained research assistants' | Yes | Yes | No | N/A |
| Bigelow et al. (2010) * | Acknowledged small sample. Acknowledged large age range of infants. Acknowledged high risk population and issues with generalizability of findings. Reported how subsample was extracted from broader trial study. Did not report income details. | No | NR | NR | Yes | No | N/A |

| Study | Reporting of sample bias | Coder training reported | Type of coder | Competing demands | Reported IRR/ICC | Reported temporality | Instrument used to measure contextual indicator |
|---|---|---|---|---|---|---|---|
| Bordeleau, Bernier & Carrier (2012) | Acknowledged relatively small sample size. Specified that birth list recruitment was random. | No | 'Research assistants' | Yes | Yes | No | N/A |
| Bouvette-Turcot, Bernier & Leblanc (2017) * | Acknowledged low risk of sample. Specified that birth list recruitment was random. | Yes | 'Trained research assistants' | Yes | Yes | Yes, No | Parenting Stress Index; Dyadic Adjustment Scale; Psychiatric Symptoms Index |
| Brooker & Poulin-Dubois (2013) * | Acknowledged recruitment from affluent, non-clinical population and issues with generalizing to more diverse population. Did not specifically acknowledge relatively small sample size. | No | 'Trained coders' | NR | Yes | Yes | N/A |
| Dayton, Huth-Bocks & Busuito (2016) | Acknowledged that sample is subsample of larger longitudinal study but did not specifically acknowledge basis of subsample eligibility. Clearly acknowledged risk status of sample with supporting demographic data. Provided details of attrition. | Yes | 'Research assistants' | Yes | No | Yes | Marital Relationship Scale; Conflict Tactics Scale-2 |
| Deschenes et al. (2014) | Acknowledged modest sample size and weakened statistical power. Acknowledged that mother-child and father-child dyads were assessed in different settings, with implications for role of SES etc. Specified that birth list recruitment was random. | No | 'Trained assistants' | Yes | No | Yes | N/A |

| Study | Reporting of sample bias | Coder training reported | Type of coder | Competing demands | Reported IRR/ICC | Reported temporality | Instrument used to measure contextual indicator |
|---|---|---|---|---|---|---|---|
| Khoury et al. (2016) * | Did not provide specific details of sample recruitment from the community but provided links to other publications. Acknowledged limitations and generalisability issues based on low-risk community sample. | No | 'Female experimenters' | NR | Yes | Yes | Beck Depression Inventory-II |
| Lemelin, Tarabulsy & Provost (2006) | Acknowledged relatively small sample size and suggested study replication. Did not provide details/criteria of participant selection process from hospital (possible selection bias). High-risk and low-risk groups demonstrated to be significantly different on sociodemographic variables. | Yes | 'Trained home visitors' | Yes | Yes | No | N/A |
| Logsdon et al. (2015) * | Acknowledged relatively small sample size. MBQS coded from very short observation time but limitation of this not acknowledged. Did not report income details. | No | 'Research assistants with expertise and extensive training' | NR | No | Yes | Clinical diagnosis of depression provided by research psychiatrists (Hamilton Scale) |
| Moran et al. (1992) * | Acknowledged small sample size. Reported criteria for selection of subsample. Relatively low mean sensitivity is explained with reference to unique risk sample type. Did not report income details. | No | 'Faculty and graduate students in developmental psychology' | Yes | Yes | Yes | N/A |
| Pederson et al. (1990) * | Did not overtly acknowledge very small sample size. Did not report possible bias associated with volunteer subject pool. Acknowledged observer's possible social perception bias. | N/A (protocol paper) | 'Faculty and graduate students in developmental psychology' | Yes | Yes | Yes | Parenting Stress Index |

| Study | Reporting of sample bias | Coder training reported | Type of coder | Competing demands | Reported IRR/ICC | Reported temporality | Instrument used to measure contextual indicator |
|---|---|---|---|---|---|---|---|
| Pereira et al. (2012) * | Acknowledged demographically low-risk status of sample. Acknowledged assessment of sensitivity at single time point. Acknowledged mothers-only sample. Acknowledged lack of assessment of potential stress moderators e.g., social support. | No | NR | Yes | Yes | Yes | Parenting Stress Index |
| Posada et al. (2004) * | Small sample size not specifically acknowledged but specified that study process will be replicated in larger and more diverse sample. Did not report income details but compared other sample and sensitivity characteristics with those in Western and other middle-class populations. | Yes | 'Trained observers' | NR | Yes | No | N/A |
| Rifkin-Graboi et al (2015) * | Described basis of subsample selection/eligibility from larger cohort study. Did not provide sociodemographic details. | Yes | 'Trained coders' | NR | Yes | Yes, No | NR |
| Rochette & Bernier (2014) | Acknowledged sample is demographically low-risk therefore cannot generalize findings to disadvantaged families. Did not specify whether recruitment from birth lists is random. | Yes | 'Trained research assistants' | Yes | Yes | No | N/A |
| Tarabulsy et al. (1997) | Acknowledged that due to the nature of observer training, it would be beneficial to develop criteria to help determine the validity of home observations. Reported statistical differences between sample sub-groups. | No | 'Observers' | Yes | No | Yes | Parenting Stress Index |

| Study | Reporting of sample bias | Coder training reported | Type of coder | Competing demands | Reported IRR/ICC | Reported temporality | Instrument used to measure contextual indicator |
|---|---|---|---|---|---|---|---|
| Tarabulsy et al. (2005) * | Acknowledged relatively small sample size. Acknowledged possible impact of other, unmeasured, ecological variables. Discussed findings with reference to nature of high-risk adolescent sample status. | Yes | 'Trained observers' | Yes | Yes | Yes | Self-report depressive symptoms (CES-D) |
| Tarabulsy et al. (2008) * | Discussed findings with reference to nature of high-risk adolescent sample status. Reported that replication of findings is necessary to further understand specific sample nuances. | Yes | 'Trained observers' | NR | Yes | No | N/A |
| Wade et al. (2015) * | Did not acknowledge limitation of sensitivity coded on shorter observation time than many studies. Described in-home data collection as limitation (with less control than lab context). | No | 'Extensively trained coders' | Yes | Yes | Yes | NR |
| Wen et al. (2017) | Described basis of subsample selection/eligibility from larger cohort study. Did not provide sociodemographic details. | Yes | 'Southeast Asian coders' | NR | Yes | Yes | N/A |
| Whipple, Bernier & Mageau (2011) | Suggested replication in larger and more socioeconomically diverse samples. Specified that recruitment from birth lists was random. | Yes | 'Trained observers' | Yes | Yes | Yes | N/A |
| Whipple, Bernier & Mageau (2011) (2) | Acknowledged low-risk status of sample with reference to greater stability of sensitivity over time, with implications for gap in sensitivity and attachment assessments. Did not report income mean. Specified that recruitment from birth lists was random. | Yes | 'Trained observers' | Yes | Yes | Yes | N/A |

| Study | Reporting of sample bias | Coder training reported | Type of coder | Competing demands | Reported IRR/ICC | Reported temporality | Instrument used to measure contextual indicator |
|---|---|---|---|---|---|---|---|
| Xing et al. (2017) * | Acknowledged modest sample size. Acknowledged limited socioeconomic diversity. | No | ‘Trained graduate students in psychology’ | NR | Yes | Yes | N/A |
| Zreik, Oppenheim & Sagi-Schwartz (2017) * | Acknowledged rationale of sample selection in specific demography cities. Acknowledged majority middle class sample. | Yes | ‘Trained observers’ | Yes | Yes | No | N/A |

*Notes.* NR = Not Reported; Reporting of sample bias =study sufficiently acknowledged sample bias/study limitations relevant to sensitivity/stress findings if applicable; Coder training reported=study reported details of MBQS coder training procedure; Competing demands=competing demands/divided attention component included in observation for MBQS coding; Reported temporality=reported measurement temporality of stress indicator/s in relation to MBQS assessment; Instrument used to measure contextual indicator=assessment of contextual indicator/s (N/A if sociodemographic only).
*Included for meta-analysis

Table 6
Summary of meta-analytic findings for effect sizes, heterogeneity and publication bias (*N*=30 effects).

| Theme | *N* effects | Combined *N* of participants | *r* (95% CI) | *p* value for *r* | SE (*r*) | $I^2$ (95% CI) | Eggar’s test *p* |
|---|---|---|---|---|---|---|---|
| Maternal education | 9 | 1059 | 0.26 (0.18, 0.34) | <.0001 | 0.04 | 38.39 (0, 84.54) | 0.29** |
| Income | 4 | 490 | 0.29 (0.19, 0.39) | <.0001 | 0.05 | 27.98 (0, 92.38) | 0.66** |
| Composite SES | 2 | 223 | 0.29 (0.17, 0.41) | <.0001 | 0.06 | 0 (0, 99.56)* | N/A |
| [a]Combined: maternal education; family income; composite SES | 15 | 1324 | 0.26 (0.19, 0.33) | 0.0002 | 0.03 | 0 (N/A)* | N/A |
| Maternal age | 4 | 264 | 0.32 (0.16, 0.49) | 0.0001 | 0.08 | 41.55 (0, 91.33) | 0.77** |

| Theme | *N* effects | Combined *N* of participants | *r* (95% CI) | *p* value for *r* | SE (*r*) | $I^2$ (95% CI) | Eggar's test *p* |
|---|---|---|---|---|---|---|---|
| Absence of father from home | 2 | 391 | -0.12 (-0.21, -0.02) | 0.019 | 0.05 | 0 (0, 99.08)* | N/A |
| Maternal internalizing symptoms | 7 | 1239 | -0.12 (-0.18, -0.05) | 0.0005 | 0.03 | 25.01 (0, 85.63) | 0.70** |
| Parenting stress | 2 | 331 | -0.13 (-0.23, -0.02) | 0.019 | 0.05 | 0 (0, 96.89)* | N/A |

*Note*. A meta-analysis was not conducted if the total number of associations available <2. N = total number; r = mean effect size; CI = confidence interval; $I^2$ = indicator of heterogeneity in percentages.
*An $I^2$ value of 0 suggests no heterogeneity in effect sizes across studies, but as these values are surrounded by large confidence intervals we cannot accurately establish the extent of heterogeneity for these groups of studies.
**Eggar's test results reported here are not accurately interpretable based on small numbers of eligible studies.
[a]Based on assumed within cluster correlation of rho=.8. Notably, sensitivity analysis suggested that the effect was robust to a broad range of assumed within cluster correlations.

Table 7
Results of sensitivity analyses.

| Theme | *r* (95% CI) | Reported unadjusted effect size ** | Stress indicator measured as antecedent to MBQS | Coded from MBQS observation time >20 minutes | Reported MBQS IRR >.80 | Used MBQS-90 (long form) | Did not publish on targeted risk sample | Mean infant age > 6 months |
|---|---|---|---|---|---|---|---|---|
| Maternal education | 0.26 (0.18, 0.34) | N/A | N/A | 0.31 (0.22, 0.40) | 0.24 (0.15, 0.32) | 0.30 (0.19, 0.41) | 0.24 (0.15, 0.32) | 0.32 (0.24, 0.41) |
| Income | 0.29 (0.19, 0.39) | N/A | 0.28 (0.13, 0.43) | 0.37 (0.25, 0.50) | 0.27 (0.16, 0.39) | N/A | N/A | 0.35 (0.24, 0.47) |
| Composite SES | 0.29 (0.17, 0.41) | N/A | N/A | N/A | N/A | N/A | N/A | N/A |

| Theme | *r* (95% CI) | Reported unadjusted effect size ** | Stress indicator measured as antecedent to MBQS | Coded from MBQS observation time >20 minutes | Reported MBQS IRR >.80 | Used MBQS-90 (long form) | Did not publish on targeted risk sample | Mean infant age > 6 months |
|---|---|---|---|---|---|---|---|---|
| Combined: maternal education; family income; composite SES | 0.26 (0.19, 0.33) | N/A | 0.26 (-0.47, 0.99) | 0.33 (0.25, 0.40) | 0.25 (0.18, 0.32) | 0.31 (0.23, 0.39) | 0.25 (0.18, 0.32) | 0.31 (0.26, 0.37) |
| Maternal age | 0.32 (0.16, 0.49) | N/A | N/A | N/A | N/A | N/A | N/A | N/A |
| Absence of father from home | -0.12 (-0.21, -0.02) | N/A | N/A | N/A | N/A | N/A | N/A | N/A |
| Maternal internalizing symptoms* | -0.12 (-0.18, -0.05) | -0.12 (-0.19, -0.04) | -0.17 (-0.30, -0.04) | -0.23 (-0.37, -0.09) | N/A | -0.11 (-0.24, 0.02) | -0.13 (-0.21, -0.05) | -0.12 (-0.21, -0.00) |
| Parenting stress | -0.13 (-0.23, -0.02) | N/A | N/A | N/A | N/A | N/A | N/A | N/A |

*Note*. X = ineligible for sensitivity analysis.
*Extra sensitivity analysis conducted for this aggregation to examine whether the effect of internalizing symptoms was robust to the inclusion of heterogeneous psychopathology types. When anxiety-related and composite symptomatology measures were removed, leaving only effects from studies that assessed pure depressive symptoms (*n*=5), the aggregate effect remained statistically significant and unchanged in magnitude (*r*=-0.10, CI=-0.19, -0.01).
**Sensitivity analyses that examined only effects representing a Pearson's *r* coefficient were identical to those reported in this column.

**Supplementary Tables**

Table S1
Search terms used for systematic review of studies using the Maternal Behavior Q-Sort.

| Search terms (PsycINFO; CINAHL; MEDLINE; Cochrane Library; SCOPUS; Google Scholar) | Search terms (EMBASE) |
|---|---|

| Search terms (PsycINFO; CINAHL; MEDLINE; Cochrane Library; SCOPUS; Google Scholar) | Search terms (EMBASE) |
|---|---|
| MBQS | MBQS |
| “maternal behavi* q-sort” | ‘maternal behavior q-sort’ |
| “maternal behavi* q sort” | ‘maternal behavior q-sort’ |
| “maternal behavi* q-set” | ‘maternal behavior q sort’ |
| “maternal behavi* q set” | ‘maternal behavior q sort’ |
| | ‘maternal behavior q-set’ |
| | ‘maternal behavior q-set’ |
| | ‘maternal behavior q set’ |
| | ‘maternal behavior q set’ |

Table S2
Risk of bias (*N*=29 studies).

| Study | Sample n >50 | MBQS observation time >20 minutes** | Reported MBQS IRR >.80** | Use of MBQS-90 (long-form)** | Report of non-adjusted stress/ sensitivity effect** | Did not use targeted risk sample** (N=targeted risk sample) | Stress antecedent** | Mean infant age > 6 months** |
|---|---|---|---|---|---|---|---|---|
| Bailey et al. (2016) * | Y | Y | Y | Y | Y | Y | N | Y |
| Bernier et al. (2010) * | Y | Y | Y | Y | Y | Y | Y, N | Y |
| Bernier & Matte-Gagne (2011) | Y | Y | Y | Y | Y | Y | N | Y |
| Bernier et al. (2014) | Y | Y | Y | Y | Y | Y | N | Y |
| Bernier, McMahon & Perrier (2017) * | Y | NR | Y | Y | Y | Y | NR | Y |
| Bigelow et al. (2010) * | N | Y | Y | Y | Y | N | N | Y |
| Bordeleau, Bernier & Carrier (2012) | Y | Y | Y | Y | Y | Y | N | Y |

| Study | Sample n >50 | MBQS observation time >20 minutes** | Reported MBQS IRR >.80** | Use of MBQS-90 (long-form)** | Report of non-adjusted stress/ sensitivity effect** | Did not use targeted risk sample** (N=targeted risk sample) | Stress antecedent** | Mean infant age > 6 months** |
|---|---|---|---|---|---|---|---|---|
| Bouvette-Turcot, Bernier & Leblanc (2017) * | Y | Y | Y | Y | Y | Y | NR | Y |
| Brooker & Poulin-Dubois (2013) * | N | N | Y | N | Y | Y | N | Y |
| Dayton, Huth-Bocks & Busuito (2016) | Y | NR | NR | N | Y | N | Y, N | Y |
| Deschenes et al. (2014) | Y | Y | NR | Y | Y | Y | N | Y |
| Khoury et al. (2016) * | Y | NR | Y | Y | Y | Y | N | Y |
| Lemelin, Tarabulsy & Provost (2006) | Y | Y | Y | Y | Y | Y, N (2 subgroups) | N | Y |
| Logsdon et al. (2015) * | Y | N | NR | NR | N*** | Y, N (2 subgroups) | Y | Y |
| Moran et al. (1992) * | N | Y | Y | Y | Y | N | N | Y |
| Pederson et al. (1990) * | N | Y | N | Y | Y | Y | N | Y |
| Pereira et al. (2012) * | Y | Y | Y | Y | Y | Y | N | Y |
| Posada et al. (2004) * | N | Y | Y | Y | Y | Y | N | Y |
| Rifkin-Graboi et al (2015) * | Y | N | Y | N | Y | Y | N | N |

| Study | Sample n >50 | MBQS observation time >20 minutes** | Reported MBQS IRR >.80** | Use of MBQS-90 (long-form)** | Report of non-adjusted stress/ sensitivity effect** | Did not use targeted risk sample** (N=targeted risk sample) | Stress antecedent** | Mean infant age > 6 months** |
|---|---|---|---|---|---|---|---|---|
| Rochette & Bernier (2014) | Y | Y | Y | Y | Y | Y | N | Y |
| Tarabulsy et al. (1997) | Y | Y | Y | Y | Y | Y, N (2 subgroups) | N | Y |
| Tarabulsy et al. (2005) * | Y | NR | Y | Y | Y | N | N | Y |
| Tarabulsy et al. (2008) * | Y | Y | Y | Y | Y | Y, N (2 subgroups) | N | Y |
| Wade et al. (2015) * | Y | N | Y | NR | Y | Y | N | N |
| Wen et al. (2017) | Y | N | Y | N | Y | Y | N | N |
| Whipple, Bernier & Mageau (2011) | Y | Y | Y | Y | Y | Y | N | Y |
| Whipple, Bernier & Mageau (2011) (2) | Y | Y | Y | Y | Y | Y | N | Y |
| Xing et al. (2017) * | Y | Y | N | N | Y | Y | N | Y |
| Zreik, Oppenheim & Sagi-Schwartz (2017) * | Y | Y | Y | N | Y | Y | NR | Y |

*Note*. Y = Yes; N = No; NR = Not Reported.
*Included for meta-analysis
**Subject to sensitivity analysis
***Converted to Pearson’s *r*